\documentclass{jpp}     

\usepackage{amsmath} 
\usepackage{amssymb}
\usepackage{import}

\usepackage[setpagesize=false,colorlinks=true,linkcolor=blue,citecolor=blue]{hyperref}  
\usepackage{cleveref}
\crefformat{section}{\S#2#1#3} 
\crefformat{subsection}{\S#2#1#3}
\crefformat{subsubsection}{\S#2#1#3}
\crefformat{figure}{figure~#2#1#3}
\crefformat{equation}{(#2#1#3)}
\crefrangeformat{equation}{(#3#1#4)--(#5#2#6)}
\crefrangeformat{figure}{figures #3#1#4--#5#2#6}
\usepackage{tikz}
\usetikzlibrary{calc}
\usepackage{pgfplots}
\pgfplotsset{compat=1.11}

\def\closesymbol{\!}

\newcommand{\rmd}{\text{d}}

\ifdefined\closesymbol
	\relax
\else
	\def\closesymbol{\relax}
\fi

\newcommand{\px}{\partial_x}
\newcommand{\py}{\partial_y}

\ifdefined\partd
	\renewcommand{\partd}[3][]{\frac{{\partial^{#1} #2}}{{\partial #3}^{#1}}}
\else
	\newcommand{\partd}[3][]{\frac{{\partial^{#1} #2}}{{\partial #3}^{#1}}}
\fi

\ifdefined\vec  
	\renewcommand{\vec}[1]{\boldsymbol{#1}}
\else
	\newcommand{\vec}[1]{\boldsymbol{#1}}
\fi

\newcommand{\uvec}[1]{\hat{\boldsymbol{#1}}}
\providecommand{\bcdot}{\vec{\cdot}}

\newcommand{\grad}{{\vec{\nabla}}}
\ifdefined\div
	\renewcommand{\div}{{\vec{nabla}}}
\else
	\newcommand{\div}{{\vec{nabla}}}
\fi

\ifdefined\vr
	\renewcommand{\vr}{{\vec{r}}}
\else
	\newcommand{\vr}{{\vec{r}}}
\fi

\newcommand{\vk}{{\vec{k}}}

\ifdefined\vv
	\renewcommand{\vv}{{\vec{v}}}
\else
	\newcommand{\vv}{{\vec{v}}}
\fi

\newcommand{\vkperp}{{\vk_\perp}}
\newcommand{\kperp}{k_\perp}
\newcommand{\kpar}{k_\parallel}

\newcommand{\intw}[1][3]{{\int \rmd^{#1} \vec{w} \ }}

\newcommand{\s}{s}

\newcommand{\vths}{v_{\text{th}\s}}
\newcommand{\vthi}{v_{\text{th}i}}
\newcommand{\vthe}{v_{\text{th}e}}

\newcommand{\rhoi}{\rho_i}
\newcommand{\rhoe}{\rho_e}

\newcommand{\fs}{f_{\s}}

\newcommand{\Fs}{F_{\s}}

\newcommand{\dfs}{\delta \closesymbol f_{\s}}

\newcommand{\Fe}{F_{e}}

\newcommand{\dfe}{\delta \closesymbol f_{e}}

\newcommand{\Fi}{F_{i}}

\newcommand{\dfi}{\delta \closesymbol f_{i}}

\newcommand{\pbra}[2]{\left\lbrace #1, #2 \right\rbrace}
\newcommand{\pbrainline}[2]{\lbrace #1, #2 \rbrace}

\newcommand{\vE}{\vec{E}}
\newcommand{\vB}{\vec{B}}
\newcommand{\exb}{\(\vE\times\vB\)}

\newcommand{\avgRs}[1]{\left\langle #1 \right\rangle_{\vec{R}_\s}}

\newcommand{\taunl}{\tau_\text{nl}}
\newcommand{\taunlo}{\tau_\text{nl}^\text{o}}
\newcommand{\phinorm}{\varphi}
\newcommand{\phinormk}{\overline{\phinorm}}
\newcommand{\phinormko}{\overline{\phinorm}^\text{o}}

\newcommand{\kxo}{k_x^\text{o}}
\newcommand{\kxotilt}{k_{x, \text{tilt}}^\text{o}}
\newcommand{\kyo}{k_y^\text{o}}
\newcommand{\gammao}{\gamma^\text{o}}

\newcommand{\taunloO}{\tau_{\text{nl}}^\text{o}(0)}
\newcommand{\kxoO}{k_{x}^\text{o}(0)}
\newcommand{\kyoO}{k_{y}^\text{o}(0)}
\newcommand{\gammaoO}{\gamma^\text{o}(0)}

\newcommand{\gammaani}{\gamma_{\text{c}}}
\newcommand{\gammaaninorm}{\hat{\gamma}_{\text{c}}}

\newcommand{\gammaE}{\gamma_{E}}
\newcommand{\gammaEhat}{\hat{\gamma}_{E}}

\newcommand{\omegaperp}{\omega_\perp}

\newcommand{\A}{\mathcal{A}}
\newcommand{\Ao}{\A^\text{o}}
\newcommand{\AoO}{\A^\text{o}(0)}

\title[]{Suppression of temperature-gradient-driven turbulence by sheared flows in fusion plasmas}

\author{P. G. Ivanov$^{1,2}$\thanks{plamen.ivanov@physics.ox.ac.uk}
	, 
	T. Adkins$^{2,3}$
	,
	D. Kennedy$^{4}$
	,
	M. Giacomin$^{5}$
	,\\
	M. Barnes$^{2,6}$
	,
	and
	A. A. Schekochihin$^{2,7}$ \\
}
\affiliation{
	$^1$Ecole Polytechnique F\'ed\'erale de Lausanne (EPFL), Swiss Plasma Center (SPC),\\CH-1015 Lausanne, Switzerland
	\\[\affilskip]
	$^2$Rudolf Peierls Centre for Theoretical Physics, University of Oxford, Oxford OX1 3PU, UK
	\\[\affilskip]
	$^3$Department of Physics, University of Otago, Dunedin, 9016, New Zealand
	\\[\affilskip]
	$^4$United Kingdom Atomic Energy Authority, Culham Campus, Abingdon OX14 3DB, UK
	\\[\affilskip]
	$^5$Dipartimento di Fisica `G. Galilei', Università degli Studi di Padova, Padova, Italy
	\\[\affilskip]
	$^6$University College, Oxford OX1 4BH, UK
	\\[\affilskip]
	$^7$Merton College, Oxford OX1 4JD, UK
}

\begin{document}
	
\maketitle

\begin{abstract}
	
\noindent Starting from the assumption that saturation of plasma turbulence driven by temperature-gradient instabilities in fusion plasmas is achieved by a local energy cascade between a long-wavelength outer scale, where energy is injected into the fluctuations, and a small-wavelength dissipation scale, where fluctuation energy is thermalised by particle collisions, we formulate a detailed phenomenological theory for the influence of perpendicular flow shear on magnetised-plasma turbulence. Our theory introduces two distinct regimes, called the weak-shear and strong-shear regimes, each with its own set of scaling laws for the scale and amplitude of the fluctuations and for the level of turbulent heat transport. We discover that the ratio of the typical radial and poloidal wavenumbers of the fluctuations (i.e., their aspect ratio) at the outer scale plays a central role in determining the dependence of the turbulent transport on the imposed flow shear. Our theoretical predictions are found to be in excellent agreement with numerical simulations of two paradigmatic models of fusion-relevant plasma turbulence: (i)~an electrostatic fluid model of slab electron-scale turbulence, and (ii)~Cyclone-base-case gyrokinetic ion-scale turbulence. Additionally, our theory envisions a potential mechanism for the suppression of electron-scale turbulence by perpendicular ion-scale flows based on the role of the aforementioned aspect ratio of the electron-scale fluctuations.
	
\end{abstract}

\section{Introduction}

The quest for controlled fusion as a viable and sustainable energy source has been a long-standing scientific and engineering challenge. The performance of magnetic-confinement-fusion devices, such as tokamaks, is often limited by the presence of turbulent fluctuations that lead to enhanced transport and energy losses. Understanding and controlling turbulence in magnetised plasmas is therefore crucial for the success of future fusion reactors. One important aspect that has attracted considerable attention is the impact of sheared flows on the turbulence \citep{artun92, synakowski97, waltz98, hobbs01, mantica09, mckee09, casson09, roach09, highcock10, barnes_shear11, field11, fedorczak13, ghim14, fox17, seiferling19}. Such sheared flows can either be externally imposed on the turbulent fluctuations as part of the plasma equilibrium, or be self-generated by the turbulence in the form of quasistatic large-scale fluctuations known as zonal flows \citep{rogers00,diamond05, dif10, dif15, zhu20_prl, zhu20_jpp, ivanov20, ivanov22}. Sheared flows can modify the size and shape of the fluctuations, and thus have a direct impact on the transport properties of the plasma.

Despite the absence of a rigorous theory of the saturation of turbulence in magnetised plasmas, it is still possible to develop phenomenological models that, at least in some regimes, capture its essential features and allow us to make falsifiable, qualitative, and sometimes even quantitative, predictions for the dependence of important turbulent properties, like the heat and particle diffusivity, on the relevant plasma parameters. Such models are often reminiscent of the original theory of hydrodynamic turbulence by \citet{K41}, which posits a local energy cascade from the outer (or injection) scale --- where energy is injected into turbulent fluctuations either by external forcing or by linear instabilities --- through the inertial range, where the nonlinear interactions dominate the dynamics and pass the energy injected at large scales down to dissipative ones \citep{GS95,sch09,barnes11,nazarenko11,adkins22,adkins23}; the energy of the fluctuations is then thermalised at these small scales, heating the plasma. The rate at which this cascade removes energy from the outer scale determines the overall turbulent amplitude and, when that is not externally imposed, the outer scale itself; in turn, the fluctuation amplitude and outer scale determine the transport. An imposed or self-generated sheared flow plays a nontrivial role in all of this.

In this article, we consider the effects of an imposed perpendicular flow shear on saturated electrostatic gyrokinetic (GK) turbulence. We first give a short recap of some relevant features of the GK framework in \cref{sec:gk}, and then, in \cref{sec:energy_cascade}, remind the reader of the standard results for saturation of such turbulence based on a local-energy-cascade phenomenology. In \cref{sec:flow_shear}, we proceed to develop a phenomenological theory of the effect of flow shear on the saturated turbulent state. The effect of this shear is to suppress the turbulent fluctuations and, in turn, the turbulent heat flux according to a certain scaling with the size of the shear. Depending on the magnitude of the imposed flow shear in comparison with the `natural' (i.e., that in the absence of shear) rate of energy injection into the fluctuations, we distinguish weak-shear (\cref{sec:weak_shear}) and strong-shear (\cref{sec:strong_shear}) regimes, each with its own scaling laws for the dependence of the turbulent transport on the shear. To verify our theoretical predictions, in \cref{sec:numerical_results}, we present numerical results from a simple electrostatic fluid model of turbulence driven by the electron-temperature-gradient (ETG) instability (\cref{sec:ETG}) and from gyrokinetic simulations of turbulence driven by the ion-temperature-gradient (ITG) instability (\cref{sec:ITG}). Then, in \cref{sec:momflux}, we discuss the transport of momentum in the electrostatic fluid model, before finally summarising and discussing our results in \cref{sec:summary}.

\section{Gyrokinetics}
\label{sec:gk}

We consider turbulent fluctuations in magnetised plasmas that satisfy the GK ordering \mbox{\(\kperp \rho_\s \sim \kpar L \sim 1\)} and \mbox{\(\omega / \Omega_\s \sim \rho_\s / L \ll 1\)}, where \(\kperp\) and \(\kpar\) are the typical perpendicular and parallel (to the mean magnetic field) wavenumbers, \(\rho_\s\) and \(\Omega_\s\) are the Larmor radius and frequency of the charged particles of species \(\s\), \(\omega\) is the inverse time scale associated with the turbulent fluctuations, and \(L\) is the length scale of variation of the plasma equilibrium. Under this ordering, we expand the distribution function for each species into equilibrium and fluctuating parts \(\fs = \Fs + \dfs\) to obtain the GK equation that governs the dynamics of the fluctuations.\footnote{Throughout this article, we use `GK' to refer to what is commonly known as `local, \(\delta\!f\) gyrokinetics'. Here `local' specifies that we are integrating \cref{eq:gk} in a domain of perpendicular size that is infinitesimal in comparison with the length scale of the variation of the plasma equilibrium, and so the gradients associated with that equilibrium are taken to be constant; `\(\delta\!f\)'~means that we only consider the evolution of small-amplitude fluctuations over times that are short compared to the transport (i.e., equilibrium-variation) time scale, with the equilibrium therefore assumed constant in time.} With the additional assumption that the plasma beta \(\beta_s \equiv 8\pi n_{\s} T_{\s} / B^2\) is small, \(n_{\s}\) and \(T_{\s}\) being the equilibrium density and temperature of species \(\s\), respectively, we can neglect the fluctuations of the magnetic field, leading to
\begin{align}
	\left(\frac{\partial}{\partial t} +  \vec{u}\bcdot\frac{\partial}{\partial \vec{R}_s}\right) \left(h_\s - \frac{q_\s \avgRs{\phi}}{T_{\s}} \Fs\right) +  w_\parallel \uvec{b} \bcdot \frac{\partial h_\s}{\partial \vec{R}_s} + \vec{v}_{d\s}& \bcdot \frac{\partial h_\s}{\partial \vec{R}_s} + \vec{v}_E \bcdot \frac{\partial}{\partial \vec{R}_s} \left(\Fs + h_s\right) \nonumber \\
	&  = \sum_{\s'}\avgRs{C_{\s \s'}}
	\label{eq:gk}
\end{align}
where the perturbed distribution function of species \(\s\) is 
\begin{equation}
	\dfs(\vec{r}, \vec{w}) = h_\s(\vec{R}_\s, \varepsilon_\s, \mu_\s) - \frac{q_\s  \phi(\vec{r})}{T_{\s}}\Fs, 
	\label{eq:delta_f}
\end{equation}
\mbox{\(\vec{R}_\s = \vec{r} - \uvec{b}\times\vec{w}/\Omega_\s\)} is the gyrocentre, \(\vec{w} = \vec{v} - \vec{u}\) is the peculiar velocity,  \mbox{\(\varepsilon_\s = m_s w^2/2\)}, \mbox{\(\mu_\s = m_s w_\perp^2 / 2B\)}, \(\phi\) is the perturbed electrostatic potential, \(\Fs\) is the equilibrium Maxwellian distribution with density \(n_{\s}\) and temperature \(T_{\s}\), \(\vec{u}\) is the equilibrium plasma flow (same for all species, see \citealt{abel13}), the magnetic drifts are \mbox{\(\vec{v}_{d\s} = (\uvec{b}/2\Omega_\s)\times(2w_\parallel^2 \uvec{b}\bcdot\grad\uvec{b} + w_\perp^2 \grad \ln B)\)}, the perturbed \exb{} drift is \mbox{\(\vec{v}_E = (c/ B) \uvec{b}\times \grad \! \avgRs{\phi} \)}, \(\uvec{b}\) is the unit vector parallel to the mean magnetic field, \(q_\s\) is the charge of species \(\s\), \(C_{\s \s'}\) is the linearised Fokker-Planck operator for collisions between particles of species \(\s\) and \(\s'\), and \(\avgRs{\dots}\) denotes the standard gyroaverage. A comprehensive derivation of the GK equations can be found in, e.g., \citet{abel13} or \citet{catto2019}. Note that the theoretical analysis presented in \cref{sec:nonlinear_saturation} does not depend on a particular coordinate system, i.e., the precise choice of radial, poloidal, and parallel coordinates, labelled \(x\), \(y\), and \(z\), respectively, will be irrelevant.

The nonlinear-advection and linear-drive terms in \cref{eq:gk} are
\begin{align}
	\vec{v}_E \bcdot \frac{\partial h_\s}{\partial \vec{R}_s} &= \frac{c}{B} \uvec{b} \bcdot (\grad x \times \grad y)  \pbra{\avgRs{\phi}}{h_\s}, \label{eq:nl_term} \\
	\vec{v}_E \bcdot \frac{\partial \Fs}{\partial \vec{R}_s} &= -\frac{c}{B} \uvec{b} \bcdot (\grad x \times \grad y)  \frac{\partial \avgRs{\phi}}{\partial y} \frac{\partial \Fs}{\partial x},
	\label{eq:drive}
\end{align}
respectively, where \(\pbrainline{f}{g} = (\px f) (\py g) - (\py f) (\px g)\). The nonlinear term \cref{eq:nl_term} expresses the advection of the perturbed distribution function by the perturbed \exb{} flow, while the linear term \cref{eq:drive} represents the injection of free energy by the radial gradients of the equilibrium (via the advection of that equilibrium by the perturbed flows). The electrostatic GK equation is closed by the quasineutrality condition:
\begin{equation}
	\sum_\s q_\s \intw \dfs = 0,
	\label{eq:qn}
\end{equation}
where the velocity integral is evaluated at fixed \(\vec{r}\).

Finally, fluctuations that evolve according to \cref{eq:gk} can be shown to satisfy a free-energy conservation law \citep{abel13} of the form
\begin{equation}
	\frac{\rmd W}{\rmd t} = I - D,
	\label{eq:W_is_I_minus_D}
\end{equation}
where the free-energy density \(W\) in a plasma of volume~\(V\) is given by
\begin{equation}
	W = \sum_\s \int \frac{\rmd^3\vr}{V} \intw \frac{T_{\s} \dfs^2}{2\Fs}.
	\label{eq:free_energy}
\end{equation}
The dissipation \(D\) in \cref{eq:W_is_I_minus_D} arises due to particle collisions and is a sink of free energy. Its precise form will not be needed here. The free-energy injection rate \(I\) depends on the gradients of the equilibrium distribution \(\Fs\) and can be written as\footnote{Strictly speaking, \cref{eq:injection_def} contains another injection term that is associated with the radial gradient of \(\vec{u}\). Here, we assume that this can be neglected (see also \cref{footnote:pvg}).}
\begin{equation}
	I = -\sum_\s \left[\Gamma_\s T_{\s} \left(\frac{\partial \ln n_{\s}}{\partial x} - \frac{3}{2}\frac{\partial \ln T_{\s}}{\partial x}\right) + Q_\s \frac{\partial\ln T_{\s}}{\partial x}\right] ,
	\label{eq:injection_def}
\end{equation}
where we have defined the flux of particles \(\Gamma_\s\) and the heat (or energy) flux \(Q_\s\) due to species \(\s\) as
\begin{align}
	\Gamma_\s \equiv &\int \frac{\rmd^3 \vec{r}}{V} \intw (\vec{v}_E \bcdot \grad x) \dfs, \\
	Q_\s \equiv &\int \frac{\rmd^3 \vec{r}}{V} \intw (\vec{v}_E \bcdot \grad x) \frac{m_\s v^2}{2} \dfs \label{eq:heatflux_def}.
\end{align}
In the most general case, \(I\) depends on both fluxes, and can be estimated as 
\begin{equation}
	I \sim \frac{\Gamma_\s T_{\s}}{L_{n_\s}}\sim \frac{Q_\s}{L_{T_\s}},
	\label{eq:injection_as_heatflux}
\end{equation}
where no additional orderings have been imposed on the density \(L_{n_\s}^{-1} \equiv - \partial_x \ln n_{\s}\) and the temperature \(L_{T_\s}^{-1} \equiv -\partial_x \ln T_{\s}\) gradients, viz., \mbox{\(L_{n_\s} \sim L_{T_\s} \sim L\)}. In this work, we consider only temperature-gradient-driven instabilities in cases where \(\Gamma_\s=0\), and so our main focus will be on \(Q_\s\). Nevertheless, the arguments presented in \cref{sec:nonlinear_saturation} are readily generalisable to cases where the injection of energy is dominated by the particle flux rather than the heat flux.

\section{Nonlinear saturation}
\label{sec:nonlinear_saturation}

Magnetised-plasma turbulence exhibits a broad range of different saturation mechanisms, and so a universal theory of turbulent saturation under the influence of flow shear is not feasible. Instead, here we focus on one particular type of saturated state, viz., that for which the following two assumptions hold: (i)~there is a scale separation between the energy-injection scale (the \textit{outer scale}), where linear instabilities inject free energy into the fluctuations, and the \textit{dissipation scale}, where fluctuations lose energy to dissipative effects (viz., ultimately, particle collisions); and (ii)~the transfer of energy between these scales is realised by a local (in scale) energy cascade. This allows us to define the \textit{inertial range} as the range of scales between the outer and dissipation scales where the local energy cascade takes place. Let us revisit the current understanding of how such an energy cascade determines the properties of the saturated turbulence.

\subsection{Outer scale, free-energy cascade, and turbulent heat flux}
\label{sec:energy_cascade}

The form of the GK nonlinearity \cref{eq:nl_term} implies that the nonlinear time \(\taunl\) at poloidal scale \(k_y\) satisfies
\begin{equation}
	\taunl^{-1} \sim \Omega_\s \rho_\s^2 k_x k_y \phinormk,
	\label{eq:taunl}
\end{equation}
where \(\phinormk\) is a measure of the characteristic amplitude of the normalised electrostatic potential \mbox{\(\varphi \equiv q_\s \phi / T_{\s}\)} at scale \(k_y\), and \(\s\) is any reference particles species. In \cref{eq:taunl}, the radial scale \(k_x\) is an implicit function of \(k_y\), viz., at each \(k_y\), the turbulent fluctuations have a typical radial scale \(k_x\) that depends on \(k_y\). With this in mind, \cref{eq:taunl} can also be written as
\begin{equation}
	\taunl^{-1} \sim \A \Omega_\s \rho_\s^2 k_y^2 \phinormk,
	\label{eq:taunl_A}
\end{equation}
where we have defined the \textit{fluctuation aspect ratio} at scale \(k_y\) as \(\A \equiv k_x/k_y\). This aspect ratio will play a critical role in the theory of sheared turbulence laid out in \cref{sec:flow_shear}. Note that the precise definition of \(\phinormk\) is not important because the phenomenological theory that is to follow predicts only scalings; nevertheless, to make things specific, one possible such definition~is
\begin{equation}
	\phinormk^2 \equiv \int_{|k_y'|>k_y} \rmd k_y' \int_{-\infty}^{+\infty} \rmd k_x'  \int \frac{\rmd z}{L_\parallel} \: |\phinorm_{\vec{k}_\perp'}|^2,
	\label{eq:phinormk_def}
\end{equation}
where \(\phinorm_\vkperp\) is the two-dimensional spatial Fourier transform of \(\varphi\) in \(x\) and \(y\), i.e., in the plane perpendicular to the equilibrium magnetic field, and \(L_\parallel\) is the parallel size of the integration domain of \cref{eq:gk} (assumed finite). By Parseval's theorem, the contributions to the free energy \cref{eq:free_energy} at each scale are proportional to the squared amplitude of the fluctuations, and so, in view of \cref{eq:delta_f} and \cref{eq:qn}, to \(\phinormk^2\).

If \mbox{\(k_x \sim k_y \sim \kperp\)} (i.e., \(\A \sim 1\)) is satisfied throughout the inertial range, \cref{eq:taunl} implies that the free-energy flux \(\varepsilon\) through scale \(\kperp\) satisfies\footnote{In general, \cref{eq:W_is_I_minus_D} implies that \(\varepsilon \sim I\). However, the free-energy flux need not equal the injection rate exactly. One such example is the fluid model of ETG turbulence described in \cref{sec:ETG}, whose free-energy-injection mechanism relies on finite collisional dissipation due to the nature of the linear instability and thus a certain order-unity fraction of \(I\) is directly dissipated at the outer scale \citep{adkins23}.}
\begin{equation}
	\varepsilon \sim \taunl^{-1} n_{\s} T_{\s} \phinormk^2 \propto \kperp^2 \phinormk^3,
	\label{eq:constant_energy_cascade}
\end{equation}
where \(\rmd/\rmd t \sim \taunl^{-1}\) in the inertial range because the dynamics there are dominated by the nonlinear effects. Assuming a local, constant-flux cascade, viz., that \(\varepsilon\) is constant throughout the inertial range, and using \cref{eq:taunl}, we conclude that
\begin{equation}
	\phinormk \propto \kperp^{-2/3} \implies \taunl^{-1} \propto \kperp^{4/3}
	\label{eq:inertial_range_scaling}
\end{equation}
in the inertial range. The assumption that \(\A \sim  1\) in the inertial range is motivated by the fact that the nonlinearity in \cref{eq:gk} is isotropic in the perpendicular plane. However, this is not a sufficient condition for \(\A \sim 1\). For example, reduced magnetohydrodynamic (RMHD) turbulence, whose nonlinearity is also isotropic in the plane perpendicular to the mean magnetic field, is known to have fluctuations that are anisotropic in the perpendicular plane and whose anisotropy depends on the scale, thus introducing a nontrivial scale-dependent factor into the RMHD version of \cref{eq:taunl} \citep{boldyrev06,boldyrev09exact, mallet16, mallet17a, schekochihin22_mhd}. Nevertheless, assuming \(\A \sim 1\) in the inertial range is not unreasonable, and it agrees with our numerical observations reported in~\cref{sec:numerical_results}.

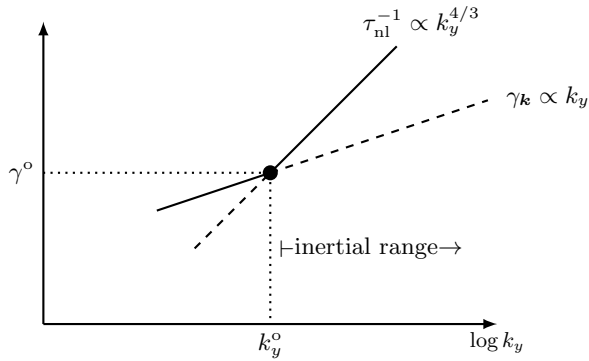
\begin{figure}
	\centering
	\scalebox{1.0}{\small
\begin{tikzpicture}[scale=1, thick, every node/.style={scale=1}]
		\coordinate (origin) at (0,0,0);    
		
		\def\xaxislength{6}
		\coordinate (xaxis) at ($ (\xaxislength,0,0) $);
		\draw[-latex] (0,0) -- (xaxis);
		\draw (xaxis) node[anchor=north,scale=0.9] {$\log{k_y}$};
		
		\def\yaxislength{4}
		\coordinate (yaxis) at ($ (0,\yaxislength,0) $);
		\draw[-latex] (0,0) -- (yaxis);
		\draw (yaxis) node[anchor=south east,scale=0.9,rotate=90] {};

		\def\yhorizontal{3.5}
		
		\def\xint{3}
		\def\yint{2}
		\node (int) at ($ (origin) + (\xint, \yint, 0) $) {};
		
		
		\def\xa{1.5}
		\def\ya{1.5}
		\coordinate (a) at ($ (origin) + (\xa, \ya, 0) $);
		
		\def\xb{6}
		\def\yb{3}
		\node (b) at ($ (origin) + (\xb, \yb, 0) $){};
		
		\node[anchor=west] at (b) {\(\gamma_\vk \propto k_y\)};
		
		\def\xc{2}
		\def\yc{1}
		\coordinate (c) at ($ (origin) + (\xc, \yc, 0) $);
		
		\def\xd{5}
		\def\yd{4}
		\node (d) at ($ (origin) + (\xd, \yd, 0) $) {\(\taunl^{-1}\propto k_y^{4/3}\)};
		
		\def\yIR{1}
		\node[anchor=west] (IRtext) at ($ (origin) + (\xint, \yIR, 0) $) {\(\mathrel{\mathop\vdash}\)inertial range\(\rightarrow\)};

		\node[anchor=north] (kyo) at ($ (origin) + (\xint, 0, 0) $) {\(\kyo\)};
		
		\node[anchor=east] (outerscale) at ($ (0, \yint, 0) $) {\(\gammao\)};
		
		\draw[-, solid, thick] (a) -- (int.center);
		\draw[-, solid, dashed] (int.center) -- (b);
		\draw[-, solid, dashed] (c) -- (int.center);
		\draw[-, solid, thick] (int.center) -- (d);
		
		\draw[-, dotted] (int) -- (kyo) {};
		\draw[-, dotted] (int) -- (outerscale) {};

		\node[circle,inner sep=2pt,fill=black] at (int) {};

	\end{tikzpicture}}
	\caption{An illustration of the relationship between the nonlinear mixing rate \(\taunl^{-1}\), the energy-injection rate \(\gamma_\vk\), and the location of the outer scale, where \(\taunl^{-1} \sim \gamma_\vk\). The scaling \(\taunl^{-1}\propto k_y^{4/3}\) is a consequence of the local energy cascade and is thus valid only in the inertial range \(k_y > \kyo\) (see the discussion in \cref{sec:energy_cascade}).}
	\label{fig:outer_scale_unsheared}
\end{figure}

Assuming that the rate of energy injection is determined by the linear-instability growth rate \(\gamma_\vk\) and that the latter satisfies \(\gamma_\vk \propto k_y\),\footnote{\label{footnote:gamma_ky}Note that, for every \(k_y\), there may be many unstable modes. For instance, in the `slab' geometry, where the linear modes are parameterised by the radial \(k_x\) and parallel \(\kpar\) wavenumbers, there exists a broad spectrum of unstable modes for any given \(k_y\). The relation \mbox{\(\gamma_\vk \propto k_y\)} does not refer to the growth rate of one particular mode (in the slab case, that would be a mode with fixed \(k_x\) and \(\kpar\)), but rather to the growth rate of a mode that is chosen by maximising the growth rate with respect to \(k_x\) and \(\kpar\) at that particular~\(k_y\). This relation is exact for long-wavelength electrostatic instabilities with \(\kperp\rho_\s \ll 1\), where the outer scale of GK turbulence often resides. In this limit, \cref{eq:gk} asymptotes to the electrostatic drift-kinetic equation, and, in the slab geometry, \(\gamma_\vk \propto k_y\) follows from the scale invariance of drift kinetics \citep{adkins23}.} the inertial-range nonlinear rate \cref{eq:inertial_range_scaling} has a steeper dependence on \(k_y\) than the injection rate. The outer scale will then be the scale at which the rates of nonlinear mixing and linear injection balance \citep{barnes11,adkins22, adkins23}:
\begin{align}
	\left(\taunlo\right)^{-1} \sim \gammao.
	\label{eq:outer_scale_def}
\end{align}
Here and in what follows, the superscript `\(^\text{o}\)' denotes quantities associated with the outer scale. The inertial range is thus located at \(k_y > \kyo\), where the nonlinear interactions dominate the linear injection rate (see \cref{fig:outer_scale_unsheared}).

\subsubsection{Heat flux}
\label{sec:heatflux}

Assuming that the heat flux \(Q_\s\) is dominated by contributions from the outer scale, we can estimate it, in view of its definition \cref{eq:heatflux_def}, as follows:
\begin{equation}
	Q_\s \sim Q^\text{o}_\s \sim  n_{\s} T_{\s} \vths \kyo \rho_\s \left(\phinormko\right)^2.
	\label{eq:heatflux_estimate}
\end{equation}
This is justified as long as the spectrum of the fluctuations decays sufficiently quickly in the inertial range; specifically, we require the fluctuation amplitudes to decay faster than $\kperp^{-1/2}$, which is readily satisfied by \cref{eq:inertial_range_scaling}. We also assume that the phase relationship between \(\phinorm\) and \(\dfs\) does not introduce any nontrivial factors --- technically, \cref{eq:heatflux_estimate} is an upper bound for~\cref{eq:heatflux_def}. Using \cref{eq:taunl_A} and \cref{eq:outer_scale_def}, we can rewrite \cref{eq:heatflux_estimate} as
\begin{equation}
	\frac{Q_\s}{n_{\s} T_{\s} \vths} \sim \left(\frac{\gammao}{\Omega_\s}\right)^2 \frac{1}{(\kyo \rho_\s)^3 (\Ao)^2}.
	\label{eq:heatflux_estimate_fancy}
\end{equation}
Therefore, in order to determine \(Q_\s\), we need to know the energy-injection rate \(\gammao\) (or equivalently \(\taunlo\)), the poloidal wavenumber \(\kyo\), and the fluctuation aspect ratio~\(\Ao\) at the outer scale. If \(\gammao\sim\gamma_{\vk^\text{o}}\), where \(\gamma_{\vk^\text{o}}\) is the growth rate at the outer scale, then only two of \(\gammao\), \(\kyo\), and \(\Ao\) are independent. Thus, we require additional assumptions. There are multiple ways to proceed.

In the absence of flow shear, \citet{barnes11} posit (i)~that the outer scale is governed by the `critical balance' of \(\gammao\) and \((\taunlo)^{-1}\) with the parallel-streaming rate across the plasma connection length, \(\omega_\parallel^\text{o} \sim \vths / qR\), where \(q\) and \(R\) are the safety factor and major radius, respectively, (ii) that the outer-scale fluctuations are isotropic, \(\Ao \sim 1\), and (iii) that the energy-injection rate is given by a simple estimate of the growth rate of temperature-gradient-driven instabilities, \mbox{\(\gammao \sim \kyo \rho_\s \vths/L_{T_\s}\)}. Combined with \cref{eq:heatflux_estimate_fancy}, \mbox{(i)--(iii)} imply
\begin{equation}
	\frac{Q_\s}{n_{\s} T_{\s} \vths} \sim \left(\frac{\rho_\s}{R}\right)^2 \Big(\frac{R}{L_{T_\s}}\Big)^3 q.
\end{equation}
Note that \citet{barnes11} studied ion-scale turbulence, which amounts to setting \(\s = i\) in the above arguments.

A modification of these results, backed by experimental \citep{ghim13} and theoretical \citep{nies24} evidence, is to replace assumption (ii) in the arguments by \citet{barnes11} by a `grand critical balance'
\begin{equation}
	\gammao \sim (\taunlo)^{-1} \sim \omega_\parallel^\text{o} \sim \omega^\text{o}_{d,x}
	\label{eq:grand_CB}
\end{equation}
between all the aforementioned rates and the radial magnetic-drift frequency \mbox{\(\omega^\text{o}_{d,x} \sim \kxo \rho_\s \vths/R\)} at the outer scale. This implies
\begin{equation}
	\Ao \sim \frac{R}{L_{T_\s}},
	\label{eq:A_grandcritbal}
\end{equation}
which, together with \cref{eq:heatflux_estimate_fancy}, results in the following scaling for the heat flux:
\begin{equation}
	\frac{Q_\s}{n_{\s} T_{\s} \vths} \sim \left(\frac{\rho_\s}{R}\right)^2 \frac{R}{L_{T_\s}} q.
\end{equation}

In the rest of this paper, we consider the influence of mean flow shear on the saturated state. We will not discuss the details of how the outer scale is determined in the case of zero imposed flow shear, but assume that the system does indeed have a well-defined zero-shear saturated state, that the outer-scale nonlinear rate is governed by \cref{eq:taunl_A} and \cref{eq:outer_scale_def}, and that \cref{eq:heatflux_estimate_fancy} is a good estimate for the heat flux. Thus, our arguments will hold regardless of whether the zero-shear outer scale is chosen \`a la \citet{barnes11}, through a grand critical balance, or otherwise.

\subsection{Perpendicular flow shear}
\label{sec:flow_shear}

For the remainder of this article, we assume an equilibrium shear flow in the direction perpendicular to the mean magnetic field and with a linear profile: \mbox{\(\vec{u} = \gammaE x \hat{\vec{y}}\)}, where \(\gammaE\) is the shearing rate.\footnote{\label{footnote:pvg}In general, a pure perpendicular linear shear is not realistic: e.g., \(\vec{u}\) is purely toroidal in axisymmetric devices, and hence has a component parallel to the mean magnetic field \citep{abel13}. For certain equilibria, the radial gradient of the parallel component of the mean flow can act as a source of energy, resulting in the so-called parallel-velocity-gradient (PVG) instability \citep{catto73,newton10,sch12}. Here we assume that there is no PVG instability (or at least that it is irrelevant for the saturated state, which is reasonable if the shear is not too large; \citealt{highcock10,barnes11}) and ignore the shear in the parallel velocity.} In the presence of such a flow, the GK equation \cref{eq:gk} is no longer homogeneous in \(x\). For brevity, we henceforth drop the species subscript from the heat flux~\(Q\).

To understand the effect of the flow shear on the fluctuations, it is instructive to consider a patch of turbulence in which the magnetic field can be considered locally constant and oriented along the $z$-direction; i.e., this patch is approximated as a `slab'. One can then perform a coordinate transformation from the original (\textit{laboratory}) frame to the so-called \textit{shearing frame} \citep{newton10, sch12}:
\begin{equation}
	t' = t, \ x' = x, \ y' = y - x\gammaE t, \ z' = z.
	\label{eq:shearing_frame}
\end{equation}
The substitution of \cref{eq:shearing_frame} into the GK equation \cref{eq:gk} eliminates the radially inhomogeneous advection term \(\vec{u}\bcdot\grad\) at the cost of introducing an inhomogeneity in time via the \(\partial_x\) derivatives. Consequently, the laboratory-frame radial wavenumber \(k_x\) of a fluctuation advected by the mean flow, i.e., of a fluctuation with a given fixed wavenumber \(\vk'\) in the shearing frame, satisfies
\begin{equation}
	k_x = k_x' - k_y' \gammaE t'.
	\label{eq:kx_drift}
\end{equation}
Crucially, the nonlinear interactions~\cref{eq:nl_term} and the linear drive~\cref{eq:drive} have the same form in both the laboratory frame and the shearing frame; therefore, \cref{eq:kx_drift} captures completely the effects of flow shear in the shearing frame. Equation \cref{eq:kx_drift} tells us that the shearing action of the perpendicular flow, which results in a `tilting' of the eddies \citep{fox17}, is equivalent to a `drift' in Fourier space of the radial wavenumber \(k_x\) of the turbulent fluctuations. 

Let us consider introducing flow shear into a system that, in its absence, would reach a saturated state by establishing a local energy cascade. As discussed in \cref{sec:energy_cascade}, the transport properties (e.g., the radial heat flux \(Q\)) of such a system are dominated by the fluctuations at the outer scale. The lifetime of these fluctuations is given by the outer-scale nonlinear time, which, according to \cref{eq:outer_scale_def}, is \(\taunloO \sim \gammaoO^{-1}\), where we will use the notation \(\gammao(\gammaE)\) to denote the dependence of outer-scale quantities on the flow shear, so \(\gammaoO\) is the outer-scale injection rate in the absence of it. We shall distinguish two different regimes of sheared turbulence: a \textit{weak-shear regime} with \mbox{\(\gammaE < \gammaoO\)} and a \textit{strong-shear regime} with \mbox{\(\gammaE > \gammaoO\)}. This distinction is motivated by the so-called `quench' rule \citep{waltz94, waltz98, kobayashi2012, ivanov20}, according to which flow shear is able to suppress the energy injection associated with some linearly unstable modes only if the shearing rate is comparable to the growth rate of those modes. If this is true, the flow shear should be unable to stifle energy injection at the outer scale if \(\gammaE < \gammaoO\), and so the outer-scale injection rate should remain independent of \(\gammaE\) in the weak-shear regime, i.e., \(\gammao(\gammaE) \approx \gammaoO\). In contrast, in the strong-shear regime, we expect that the injection rate will be modified by the presence of the flow shear.
 
Let us analyse the physics of both regimes, starting with the weak-shear one.

\subsubsection{Weak-shear regime}
\label{sec:weak_shear}

Let us consider more carefully the influence of flow shear on the outer scale in the case \(\gammaE < \gammaoO\). As just discussed, in this regime, the outer-scale-injection and nonlinear-mixing rates should remain approximately the same as those at \mbox{\(\gammaE = 0\)}, viz.,
\begin{equation}
	\taunlo(\gammaE)^{-1} \sim \gammao(\gammaE) \sim \gammaoO \sim \taunloO^{-1}.
	\label{eq:outer_scale_balance_weak_shear}
\end{equation}
Assuming that the injection rate \(\gammao\) is determined by the linear growth rate at the outer scale,\footnote{Strictly speaking, linear modes in the presence of flow shear are often only transiently growing, so \(\gammao(\gammaE)\) should be interpreted as the transient growth rate at early times \citep{highcock11, sch12}. We also note that a shear in the equilibrium magnetic field can introduce nontrivial effects, e.g., travelling exponentially growing modes \citep{newton10} or nonexponentially growing Floquet modes \citep{cooper88,waelbroeck91}, which can have a significant impact on the saturation, e.g., they may lead to a bistable saturated state \citep{christen22}. Our analysis depends only on the injection rate being a function solely of the poloidal wavenumber, while the precise relationship between the injection rate and the linear growth rate is outside of the scope of the current work.} and that the latter is (at least approximately) only a function of \(k_y\),\footnote{This assumption can be made weaker: we will only need \(\gamma_\vk\) to be approximately independent of \(k_x\) for \(k_x < k_y\).} we conclude that the poloidal wavenumber of the outer-scale eddies is also set by its value at \(\gammaE=0\) and independent of \(\gammaE\) in the weak-shear regime, viz.,
\begin{equation}
	\kyo(\gammaE) \sim \kyoO.
	\label{eq:kyo_weak_shear}
\end{equation}
However, the assumption that \(\gamma_\vk\) is only a weak function of \(k_x\) means that one cannot make a similar statement about the radial wavenumber \(\kxo(\gammaE)\). Indeed, approximating the lifetime of the outer-scale fluctuations as equal to the nonlinear mixing time \(\taunlo\), the wavenumber drift \cref{eq:kx_drift}, together with \cref{eq:outer_scale_balance_weak_shear} and \cref{eq:kyo_weak_shear}, suggests that
\begin{align}
	\kxo(\gammaE) &\sim \kxoO + \kyoO\taunloO\gammaE \nonumber \\
	&\sim \kxoO \left[1 + \frac{\gammaE}{\AoO \gammaoO}\right],
	\label{eq:kx_weak_shear}
\end{align}
where \(\AoO=\kxoO/\kyoO\) is the fluctuation aspect ratio at the outer scale at \(\gammaE = 0\). Therefore,
\begin{equation}
	\frac{\kxo(\gammaE)}{\kxoO}\sim \frac{\Ao(\gammaE)}{\AoO} \sim 1 + \frac{\gammaE}{\gammaani},
	\label{eq:kxo_weak_shear}
\end{equation}
where we have introduced the \textit{critical shearing rate}
\begin{align}
	\gammaani \equiv \AoO \gammaoO.
	\label{eq:gammaani_def}
\end{align}
Then, \cref{eq:heatflux_estimate_fancy} implies that the radial turbulent heat flux satisfies
\begin{equation}
	\frac{Q(\gammaE)}{Q(0)}\sim \left[\frac{\AoO}{\Ao(\gammaE)}\right]^2  \sim \frac{1}{(1+\gammaE/\gammaani)^2},
	\label{eq:heatflux_weakly_sheared}
\end{equation}
where \(Q(0)\) is the heat flux at \(\gammaE = 0\). Note that at no step leading to \cref{eq:heatflux_weakly_sheared} did we use any formulae from \cref{sec:energy_cascade} that relied on isotropy, which would otherwise have restricted us to \mbox{\(\Ao \sim 1\)}.

Expressions \cref{eq:gammaani_def} and \cref{eq:heatflux_weakly_sheared} predict that the transport properties in the weak-shear regime are determined by the ratio of the radial and poloidal wavenumbers of the outer-scale eddies, \(\AoO = \kxoO/\kyoO\). If the unsheared fluctuations have \(\AoO \sim 1\), then \cref{eq:gammaani_def} implies that \mbox{\(\gammaani \sim \gammaoO\)}. As the weak-shear regime is characterised by \mbox{\(\gammaE < \gammaoO\)}, \cref{eq:kxo_weak_shear} implies that \mbox{\(\Ao(\gammaE) \sim \AoO \sim 1\)}, and thus \(Q(\gammaE) \sim Q(0)\) throughout the weak-shear regime. In other words, if the unsheared turbulence has \(\AoO \sim 1\) at the outer scale, shearing it with any \(\gammaE < \gammaoO\) will not reduce the turbulent transport by more than an order-unity amount --- an unsurprising outcome.

However, due to the nature of the underlying linear instabilities, it is, in fact, often the case that the outer-scale eddies in temperature-gradient-driven turbulence are radially elongated. Such eddies, often called `streamers', are a well-documented feature of this type of turbulence, especially in its electron-scale variety \citep{drake88, jenko00, dorland00, jenko06, roach09, colyer17}. A turbulent state dominated by streamers satisfies \(\AoO \ll 1\), and so \(\gammaani \ll \gammaoO\). In this case, \cref{eq:kxo_weak_shear} implies that the outer-scale aspect ratio increases linearly with the flow shear due to the tilting of the eddies, viz.,
\begin{equation}
	\Ao(\gammaE) \sim \AoO + \frac{\gammaE}{\gammaoO}.
	\label{eq:A_weakshear}
\end{equation}
Furthermore, \cref{eq:heatflux_weakly_sheared} predicts that the heat flux will be suppressed if \(\gammaani \lesssim \gammaE \ll \gammaoO\). In particular, for intermediate values of the shearing rate that satisfy \(\gammaani \ll \gammaE \ll \gammaoO\), \cref{eq:A_weakshear} becomes 
\begin{equation}
	\Ao(\gammaE) \sim \frac{\gammaE}{\gammaoO},
	\label{eq:A_weakshear_intermediate}
\end{equation}
and so, by \cref{eq:heatflux_weakly_sheared},
\begin{equation}
	Q(\gammaE) \propto \gammaE^{-2}.
	\label{eq:intermediate_shear_heat_flux}
\end{equation}

If the shear is increased further, \cref{eq:A_weakshear} implies that, at the transition from the weak- to the strong-shear regime, where \mbox{\(\gammaE \sim \gammaoO\)}, the outer-scale aspect ratio is \(\Ao(\gammaE) \sim 1\), and so, by \cref{eq:heatflux_weakly_sheared}, the heat flux has been reduced by a large factor:
\begin{equation}
	\frac{Q[\gammaoO]}{Q(0)} \sim \left[\AoO\right]^2 \ll 1.
\end{equation}

A cautious reader may have spotted a potential clash between having \(\AoO \ll 1\) at the outer scale and the theory of the energy cascade laid out in \cref{sec:energy_cascade}: there, we assumed that the fluctuations in the inertial range had \(\A \sim 1\), yet the inertial range must connect to the outer scale, where now \(\A \ll 1\). There are two possible resolutions to this problem: (i) the scaling arguments presented in \cref{sec:energy_cascade} are, in fact, unchanged if the inertial range inherits the aspect ratio at the outer scale, i.e., if \(k_x/k_y\) is a scale-independent, even if numerically small, number below the outer scale, or (ii) there exists a transition region below the outer scale, wherein the dependence of \(k_x\) on \(k_y\) is faster than linear so that \(k_x\) gradually increases to match \(k_y\) at some smaller scale, below which the scalings from \cref{sec:energy_cascade} become valid. Our numerical results, presented in \cref{sec:ETG}, are consistent with option~(ii). In \cref{appendix:aniso_to_iso}, we develop a simple theory for the transition region.

Finally, let us mention that, while here we shall consider only cases where \mbox{\(\A \lesssim 1\)}, this is not necessarily satisfied in all instances of fusion-relevant turbulence. For example, the large-temperature-gradient environment of the pedestal has been shown numerically to give rise to poloidally elongated turbulent fluctuations with \(\A \gg 1\) \citep{parisi20, parisi22}. As discussed in \cref{sec:heatflux}, the `grand critical balance' \cref{eq:grand_CB} leads to poloidally elongated eddies at large temperature gradients, as per \cref{eq:A_grandcritbal}. Recent numerical and analytical work by \mbox{\citet{nies24}} suggests that such behaviour may indeed be consistent with strongly driven ion-temperature-gradient turbulence in axisymmetric toroidal geometry. Here, leaving the reader cognisant of these recent developments, we shall nevertheless focus on the case~\(\A \lesssim 1\).
 
\subsubsection{Strong-shear regime}
\label{sec:strong_shear}

\begin{figure}
	\centering
	\scalebox{1.0}{\small
\begin{tikzpicture}[scale=1, thick, every node/.style={scale=1}]
		\coordinate (origin) at (-1,0,0);    
		
		\def\xaxislength{6}
		\coordinate (xaxis) at ($ (\xaxislength,0,0) $);
		\draw[-latex] (0,0) -- (xaxis);
		\draw (xaxis) node[anchor=north,scale=0.9] {$\log{k_y}$};
		
		\def\yaxislength{4}
		\coordinate (yaxis) at ($ (0,\yaxislength,0) $);
		\draw[-latex] (0,0) -- (yaxis);
		\draw (yaxis) node[anchor=south east,scale=0.9,rotate=90] {};

		\def\yhorizontal{3.5}
		
		\def\xa{1.5}
		\def\ya{1.5}
		\coordinate (a) at ($ (origin) + (\xa, \ya, 0) $);
		
		\def\xb{6}
		\def\yb{3}
		\node (b) at ($ (origin) + (\xb, \yb, 0) $){};
		
		\node[anchor=north west] at (b) {\(\gamma_\vk \propto k_y\)};
		
		\def\xc{2}
		\def\yc{1}
		\coordinate (c) at ($ (origin) + (\xc, \yc, 0) $);

		\def\xd{5}
		\def\yd{4}
		\node[gray] (d) at ($ (origin) + (\xd, \yd, 0) $) {};
		
		\def\xcnew{3.5}
		\def\ycnew{1.5}
		\coordinate (cnew) at ($ (origin) + (\xcnew, \ycnew, 0) $);
		
		\def\xdnew{6}
		\def\ydnew{4}
		\node (dnew) at ($ (origin) + (\xdnew, \ydnew, 0) $) {\(\quad \taunl^{-1}\propto k_y^{4/3}\)};
		
		\def\xint{3}
		\def\yint{2}
		\node (int) at ($ (origin) + (\xint, \yint, 0) $) {};
		
		\def\xintnew{4.5}
		\def\yintnew{2.5}
		\node (intnew) at ($ (origin) + (\xintnew, \yintnew, 0) $) {};
		
		\def\yIR{1}
		\node[anchor=west] (IRtext) at ($ (origin) + (\xintnew, \yIR, 0) $) {\(\mathrel{\mathop\vdash}\)inertial range\(\rightarrow\)};

		\node[anchor=north, gray] (kyo) at ($ (origin) + (\xint, 0, 0) $) {\(\kyoO\)};
		\node[anchor=north] (kyonew) at ($ (origin) + (\xintnew, 0, 0) $) {\(\kyo(\gammaE)\)};
		
		\draw[-, solid, thick] (a) -- (intnew.center);
		\draw[-, solid, dashed] (intnew.center) -- (b);
		\draw[-, solid, dashed, gray] (c) -- (int.center);
		\draw[-, solid, thick, gray] (int.center) -- (d);
		\draw[-, solid, dashed] (cnew) -- (intnew.center);
		\draw[-, solid, thick] (intnew.center) -- (dnew);
		
		\draw[-, dotted, gray] (int) -- (kyo) {};
		\draw[-, dotted] (intnew) -- (kyonew) {};

		\node[circle,inner sep=2pt,fill=gray] at (int) {};
		\node[circle,inner sep=2pt,fill=black] at (intnew) {};

		\node[anchor=east] (shear) at (0, \yintnew, 0) {\(\gammao(\gammaE)\)};
		\draw[-, dotted, thick] (shear) -- (intnew);

		\node[anchor=east, gray] (gammaoO) at (0, \yint, 0) {\(\gammaoO\)};
		\draw[-, dotted, thick, gray] (gammaoO) -- (int);		 
		
		\node[label={[align=center,shift={(1,0.3)},fill=white,fill opacity=1]below:shear-suppressed\\eddies}] at (c) {};

	\end{tikzpicture}}
	\caption{A qualitative illustration, analogous to \cref{fig:outer_scale_unsheared}, of the effect of strong flow shear \(\gammaE \gg \gammaoO\), leading to the time-scale balance~\cref{eq:isotropic_outer_scale_balance} determining the outer scale. In this regime, \(\gammao(\gammaE) \sim \gammaE\).}
	\label{fig:outer_scale_sheared}
\end{figure}
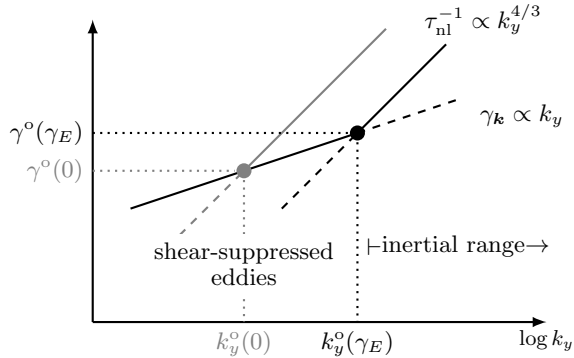

In the strong-shear regime, defined by \(\gammaE > \gammao(0)\), the flow shear is strong enough to affect energy injection at the outer scale. In particular, it is no longer possible to excite fluctuations at wavenumbers where the growth rate is \(\gamma_\vk \lesssim \gammaE\) \citep{waltz94, waltz98}. To compensate for this, the outer scale must adjust to match the shearing rate. Thus, we propose that, for \(\gammaE \gg \gammaoO\), the outer scale will be governed by the balance of nonlinear, injection, and shearing rates:
\begin{equation}
	\taunlo(\gammaE)^{-1} \sim \gammao(\gammaE) \sim \gammaE, 
	\label{eq:isotropic_outer_scale_balance}
\end{equation}
as illustrated in \cref{fig:outer_scale_sheared}. As always, the lifetime of turbulent eddies at the outer scale is set by the nonlinear time; \cref{eq:kx_drift} then implies that \(\Ao(\gammaE) \sim 1\) throughout the strong-shear regime. Assuming that the linear growth rate is \(\gamma_\vk \propto k_y\), we expect that 
\begin{equation}
	\kyo \propto \gammaE. 
	\label{eq:kyo_strong_shear}
\end{equation}
This is intuitively clear: stronger flow shear pushes turbulence towards smaller (and thus faster) scales since the larger (and slower) eddies are more strongly affected by the shear. Consequently, \cref{eq:heatflux_estimate_fancy}, together with \cref{eq:isotropic_outer_scale_balance} and \cref{eq:kyo_strong_shear}, implies
\begin{align}
	Q(\gammaE) \propto \gammaE^{-1}
	\label{eq:heatflux_strong_shear}
\end{align}
in the strong-shear regime \(\gammaE > \gammaoO\).

\begin{figure}
	\centering
	\begin{tikzpicture}[scale=1, thick, every node/.style={scale=1}]
    \coordinate (origin) at (0,0,0);
    
    \def\xaxislength{4.8}
    \coordinate (xaxis) at ($ (origin) + (\xaxislength,0,0) $);
    \draw[-latex] (0,0) -- (xaxis);
    \draw (xaxis) node[anchor=north] {$\log{\gammaE}$};
    
    \def\yaxislength{4}
    \coordinate (yaxis) at ($ (origin) + (0,\yaxislength,0) $);
    \draw[-latex] (0,0) -- (yaxis);
    \draw (yaxis) node[anchor=east] {$\log{Q}$};
    
    \def\fluxstyle{solid}
    \def\fluxthickness{thick}
    \def\vertstyle{dotted}

    \def\yhorizontal{3.5}
    
    \def\ya{\yhorizontal}
    \def\xa{0}
    \coordinate (a) at ($ (origin) + (\xa, \ya, 0) $);
    \node[anchor=east] at (a) {\(Q(0)\)};
    
    \def\yb{\yhorizontal}
    \def\xb{1.5}
    \coordinate (b) at ($ (origin) + (\xb, \yb, 0) $);
    
    \def\yc{\yhorizontal-1.5}
    \def\xc{2.5}
    \coordinate (c) at ($ (origin) + (\xc, \yc, 0) $);
    
    \def\yd{\yhorizontal-2}
    \def\xd{3.75}
    \coordinate (d) at ($ (origin) + (\xd, \yd, 0) $);
    
    \def\ye{0.0}
    \def\xe{4.75}
    \coordinate (e) at ($ (origin) + (\xe, \ye, 0) $);
    
    \coordinate (QgammaoO) at (0, \yc, 0); 
    
    \def\gammamaxx{4}
    \node[anchor=north] at ($ (origin) + (\gammamaxx, 0, 0) $) {\(\gamma_\text{max}\)};
    
    
    \node[anchor=west, rotate=0] at ($(b) - (0.1, 0.2, 0)$) {\(Q \propto (1 + \gammaE/\gammaani)^{-2}\)};
    \node[circle,inner sep=2pt,fill=black] at (c) {};
    
    \draw[-, \fluxstyle, \fluxthickness] (c) -- (d);
    \node[anchor=west, rotate=0] at ($(c) + (0.5, 0, 0)$)  {\(Q \propto \gammaE^{-1}\)};

	\draw [black,rounded corners=5ex] (d) -- ($(d) + (0.8, -0.4, 0)$) -- (e);
     
    \node[anchor=north] (gammaani) at (0.6*\xb, 0, 0) {\(\gammaani\)};
    \draw[-, \vertstyle] (gammaani) -- (0.6*\xb, 0.94*\yb, 0);    
    
    \node[anchor=north] (gammaoO) at (\xc, 0, 0) {\(\gammaoO\)}; 
    \draw[-, \vertstyle] (gammaoO) -- (c);    
    \draw[-, \vertstyle] (QgammaoO) -- (c);
    
    \draw[<->, >=stealth, \vertstyle] ($ (QgammaoO) + (0.2, 0, 0) $) -- node[midway, right, scale=1,fill=white, anchor=north west] {
    	$\begin{aligned}
    		\propto \left[\AoO\right]^2
    	\end{aligned}$
    } ($ (a) + (0.2, 0, 0) $);
    
    \draw [black,rounded corners=10ex] (a) -- (b) -- (c);
    
    \node[anchor=south] at ({0.5*(\xa+\xc)}, 0, 0) {(i)};
    \node[anchor=south] at ({0.5*(\xc+\gammamaxx)}, 0, 0) {(ii)};

	\draw[dotted, \fluxthickness] (\gammamaxx, 0, 0) -- (\gammamaxx, 0.33*\yaxislength, 0);

	\def\ysp{4.2};
	\def\xsp{-0.95};
    \node[anchor=south] at ($ (origin) + (\xsp, \ysp, 0) $) {(a)};
\end{tikzpicture}
	\begin{tikzpicture}[scale=1, thick, every node/.style={scale=1}]
	\coordinate (origin) at (0,0,0);    
	
	\def\xaxislength{4.8}
	\coordinate (xaxis) at ($ (origin) + (\xaxislength,0,0) $);
	\draw[-latex] (0,0) -- (xaxis);
	\draw (xaxis) node[anchor=north] {$\log{\gammaE}$};
	
	\def\yaxislength{4}
	\coordinate (yaxis) at ($ (origin) + (0,\yaxislength,0) $);
	\draw[-latex] (0,0) -- (yaxis);
	\draw (yaxis) node[anchor=east] {$\log{Q}$};
	
	\def\fluxstyle{solid}
	\def\fluxthickness{thick}
	\def\vertstyle{dotted}
	
	\def\yhorizontal{6}
	
	\def\ya{\yhorizontal-2.5}
	\def\xa{0}
	\coordinate (a) at ($ (origin) + (\xa, \ya, 0) $);
	\node[anchor=east] at (a) {\(Q(0)\)};
	
	\def\yb{\yhorizontal-2.5}
	\def\xb{1.5}
	\coordinate (b) at ($ (origin) + (\xb, \yb, 0) $);
	
	\def\yc{\yhorizontal-2.5}
	\def\xc{2.5}
	\coordinate (c) at ($ (origin) + (\xc, \yc, 0) $);
	
	\def\yd{2.75}
	\def\xd{3.75}
	\coordinate (d) at ($ (origin) + (\xd, \yd, 0) $);
	
	\def\ye{0.0}
	\def\xe{4.75}
	\coordinate (e) at ($ (origin) + (\xe, \ye, 0) $);
	
	\coordinate (QgammaoO) at (0, \yc, 0); 
	
	\def\gammamaxx{4}
	\node[anchor=north] at ($ (origin) + (\gammamaxx, 0, 0) $) {\(\gamma_\text{max}\)};
	
	\draw [black] (a) -- (b) -- (c);
	\node[circle,inner sep=2pt,fill=black] at (c) {};
	
	\draw[-, \fluxstyle, \fluxthickness] (c) --	(d);
	\node[anchor=west, rotate=0] at ($(c) - (-0.7, 0.25, 0)$)  {\(Q \propto \gammaE^{-1}\)};
	
	\draw [black,rounded corners=5ex] (d) -- ($(d) + (0.8, -0.5, 0)$) -- (e);
	
	\node[anchor=north] (gammaoO) at (\xc, 0, 0) {\(\gammaani \sim \gammaoO\)}; 
	\draw[-, \vertstyle] (gammaoO) -- (c);    
	
	\node[anchor=south] at ({0.5*(\xa+\xc)}, 0, 0) {(i)};
	\node[anchor=south] at ({0.5*(\xc+\gammamaxx)}, 0, 0) {(ii)};
	\draw[dotted, \fluxthickness] (\gammamaxx, 0, 0) -- (\gammamaxx, 0.62*\yaxislength, 0);
	
	\def\ysp{4.2};
	\def\xsp{-0.95};
	\node[anchor=south] at ($ (origin) + (\xsp, \ysp, 0) $) {(b)};
\end{tikzpicture}
	\caption{A qualitative diagram of the heat flux \(Q\) as a function of the flow shear \(\gammaE\) in the case of (a) \(\AoO \ll 1\) and (b)~\(\AoO \sim 1\). In each case, there are two distinct regimes. (i) For \(\gammaE < \gammaoO\), we have the weak-shear regime (\cref{sec:weak_shear}), where, in the \(\AoO \ll 1\) case, we find \(Q(\gammaE) \propto (1 + \gammaE/\gammaani)^{-2}\) [see \cref{eq:heatflux_weakly_sheared}]. In contrast, if \(\AoO \sim 1\), the flow shear is unable to affect significantly the fluctuations at the outer scale and, consequently, the heat flux is approximately independent of \(\gammaE\). (ii) For \(\gammaoO < \gammaE < \gamma_\text{max}\), the system is in the strong-shear regime (\cref{sec:strong_shear}), where the outer-scale injection rate is determined by the flow shear, viz., \(\gammao(\gammaE) \sim \gammaE\). Here \(\Ao(\gammaE) \sim 1\) at the outer scale, regardless of \(\AoO\), and \(Q(\gammaE)\propto\gammaE^{-1}\). Finally, the fluctuations, and hence the heat flux, are completely suppressed at \mbox{\(\gammaE \gtrsim \gamma_\text{max}\)}. }
	\label{fig:heatflux_anisotropic}
\end{figure}
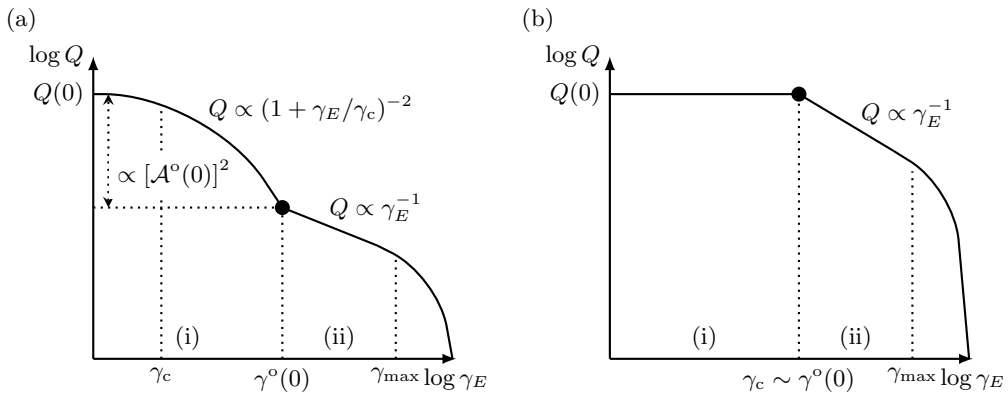

The outer-scale balance \cref{eq:isotropic_outer_scale_balance}, and thus the scaling \cref{eq:heatflux_strong_shear}, cannot be satisfied for arbitrarily large values of flow shear because the linear growth rate \(\gamma_\vk\) is bounded by some \(\gamma_\text{max}\), normally found at much larger wavenumbers than those associated with the dominant energy injection.\footnote{The existence of such \(\gamma_\text{max}\) can be proven rigorously in some cases, e.g., ion-temperature-gradient-driven turbulence with adiabatic electrons \citep{helander22}.} For \(\gammaE \gtrsim \gamma_\text{max}\), the system is no longer able to sustain the turbulent fluctuations because the shearing rate \(\gammaE\) cannot be matched by the rate of energy injection at \textit{any} scale. Therefore, we expect a sharp cutoff in the fluctuations' amplitude, and thus in the heat flux, as \(\gammaE\) becomes comparable to \(\gamma_\text{max}\). \Cref{fig:heatflux_anisotropic} summarises the expected dependence of the heat flux on \(\gammaE\) in both regimes.

\section{Numerical results}
\label{sec:numerical_results}

To test the validity of the theory presented in \cref{sec:flow_shear}, we consider two different models of turbulence. The first (\cref{sec:ETG}) is a two-fluid model that captures the dynamics of electrostatic fluctuations of density and electron temperature in a straight magnetic field. This turbulence is driven by the collisional slab ETG (sETG) instability \citep{adkins22} on scales between the ion and electron gyroradii. While this model is extremely simple, even simplistic, the benefit of using it is that its saturation mechanism has already been investigated in great detail and has been shown to conform to the picture of a local energy cascade outlined in \cref{sec:energy_cascade} \citep{adkins23}. Therefore, it is a prime candidate for confirming the validity of the theory laid out in this paper. For our second set of simulations (\cref{sec:ITG}), we employ the GK code \texttt{GENE} \citep{jenko00, jenko00GENE} to perform gyrokinetic flux-tube simulations of ITG-driven turbulence. This is a much more realistic model of plasma turbulence, and the `Cyclone base case' used here is a setup that has been extensively studied in the literature (\citealt{lin99,dimits00, barnes09, highcock12, peeters16, li21, ajay21, volcokas22, hoffmann23,lippert23,tirkas23} constitute a small sample) --- it is thus a natural testbed for any theory aspiring to tokamak relevance.

\subsection{Fluid ETG turbulence}
\label{sec:ETG}

In this section, we report numerical simulations in a triply periodic domain of size \(L_x\), \(L_y\), and \(L_\parallel\) in \(x\), \(y\), and \(z\), respectively, of the following collisional slab ETG model~\citep{adkins23}:
\begin{align}
	&\frac{\rmd}{\rmd t} \frac{\delta n_e}{n_{e}} + \frac{\partial u_{\parallel e}}{\partial z} = 0, \label{eq:phi} \\
	&\frac{\nu_{ei}}{c_1}\frac{u_{\parallel e}}{\vthe} = -\frac{\vthe}{2} \frac{\partial}{\partial z}\left[ \frac{\delta n_e}{n_{e}} - \varphi + \left(1 + \frac{c_2}{c_1}\right)\frac{\delta T_e}{T_{e}} \right], \label{eq:u}\\
	&\frac{\rmd}{\rmd t}\frac{\delta T_e}{T_{e}} - \frac{c_3\vthe^2}{3\nu_{ei}}\frac{\partial^2}{\partial z^2} \frac{\delta T_e}{T_{e}} + \frac{2}{3} \left(1 + \frac{c_2}{c_1}\right)\frac{\partial u_{\parallel e}}{\partial z} = -\frac{\rhoe \vthe}{2 L_T}\frac{\partial \varphi}{\partial y}, 
	\label{eq:T}
\end{align}
where the `convective derivative'
\begin{equation}
	\frac{\rmd}{\rmd t} = \frac{\partial}{\partial t} + \gammaE x \frac{\partial}{\partial y} + \frac{\rhoe\vthe}{2} \left(\hat{\vec{z}} \times \grad \varphi\right)\bcdot\grad + \nu_\perp \rhoe^4 \nabla_\perp^4
\end{equation}
includes the mean flow shear, the nonlinear advection by the perturbed \mbox{\exb{}} drift, and hyperviscous dissipation. \Cref{appendix:fluid_numerics} describes some important details of the numerical implementation of the flow-shearing term \(\gammaE x \partial_y\). The electron density is related to the electrostatic potential $\phinorm \equiv e \phi / T_{e}$ by quasineutrality \cref{eq:qn} combined with the assumption of adiabatic ions:
\begin{align}
	\frac{\delta n_e}{n_{e}} = - \frac{Z T_{e}}{T_{i}} \phinorm,
	\label{eq:qn_etg}
\end{align}
where \(Z = q_i / e\), \(q_i\) being the ion charge. The numerical coefficients \(c_1\), \(c_2\), and \(c_3\) arise from the physics of collisions and depend on \(Z\): e.g., for \(Z = 1\), \(c_1 \approx 1.94\), \(c_2 \approx 1.39\), and \(c_3\approx3.16\), in agreement with \citep{braginskii65}. We used \(Z = 1\) and \(T_{i} = T_{e}\) for all simulations reported here. Finally, the electron heat flux \cref{eq:heatflux_def} can be expressed using the Fourier amplitudes of the fluctuations as follows:
\begin{equation}
	Q = \frac{3}{2} n_{e}T_{e}\vthe \sum_\vk i k_y\rho_e \varphi_\vk^* \frac{\delta T_{e,\vk}}{T_{e}}.
\end{equation}

\begin{table}
	\centering
	\begin{tabular}{l  c  c  c  c  c  c}
		\vspace{1.5mm}
		& $L_\perp/\rho_\perp$ & $L_\parallel \sqrt{\sigma}/L_T$ & $n_\perp$ & $n_\parallel$ & $\nu_\perp \rhoe^4 / \omegaperp\rho_\perp^4$ & \(\hat{\gamma}_\text{max}\) \\
		Sim1 & 100 & 50 & 341 & 31 & \(5 \times 10^{-4}\) & \(6.3\times 10^2\) \\
		Sim2 & 100 & 50 & 683 & 43 & \(1 \times 10^{-4}\) & \(1.1\times 10^3\) \\
		Sim3 & 70 & 40 & 191 & 31 & \(5 \times 10^{-4}\) & \(4.1\times 10^2\) \\
		Sim4 & 40 & 30 & 191 & 31 & \(5 \times 10^{-5}\) & \(4.9\times 10^2\) \\
	\end{tabular}
	\caption{A summary of the simulation parameters used in \cref{sec:ETG}. The simulation domain is taken to be `square' with \mbox{\(L_x = L_y = L_\perp\)} and \(n_x = n_y = n_\perp\), where \(n_x\), \(n_y\), and \(n_\parallel\) are the number of resolved (i.e., after dealiasing --- see \cref{appendix:fluid_numerics}) Fourier modes in the \(x\), \(y\), and \(z\) coordinates, respectively. The last column shows the maximum growth rate \(\gamma_\text{max}\) normalised according to~\cref{eq:gammanorm}. }
	\label{tab:sims}
\end{table}

\begin{figure*}
	\begingroup\import{figs/fig4}{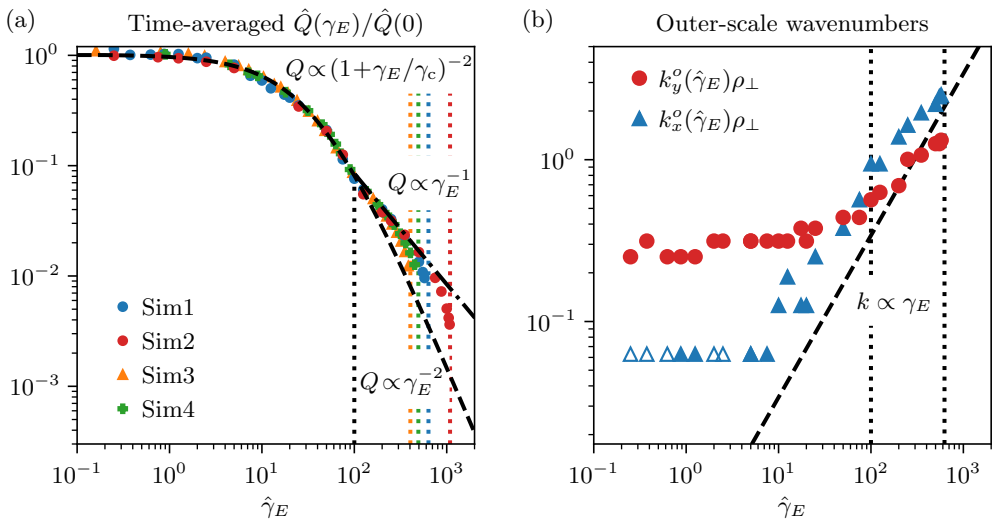}\endgroup
	\caption{(a) Time-averaged, saturated radial turbulent heat flux, normalised to its value at zero flow shear, as a function of normalised flow shear~\(\gammaEhat\) [normalised per \cref{eq:gammanorm}] for the sets of simulations detailed in \cref{tab:sims}. The data from all four sets overlays due to the scale invariance of \mbox{\cref{eq:phi}--\cref{eq:T}}. The black dashed and dash-dotted lines show the theoretical predictions \cref{eq:heatflux_weakly_sheared} and \cref{eq:heatflux_strong_shear}, respectively, where, for the former, the curve is plotted using \(\gammaaninorm \approx 39\), found by fitting to the data presented here. The vertical black dotted line marks the approximate shearing rate \(\gammaEhat\approx 100\) where the system transitions from the weak- to the strong-shear regime. The values of \(\gammaE \approx \gamma_\text{max}\) are shown using vertical dotted lines of the same colour as the data points for each respective set of simulations. (b) The outer-scale wavenumbers \(\kxo(\gammaE)\) and \(\kyo(\gammaE)\), defined as those that maximise \cref{eq:Q_poloidal_avg} and \cref{eq:Q_radial_avg}, respectively, for the Sim1 set of simulations. The dashed line indicates a linear dependence on the flow shear, \(k \propto \gammaE\). The left vertical dotted line is the same as in panel (a) and marks the location \(\gammaE \approx 100\) where the system transitions from the weak- to the strong-shear regime. In the former, \(\kyo\) is (approximately) pinned to \(\kyoO\) but \(\kxo\) increases linearly with \(\gammaE\). In the strong-shear regime, \(\kxo \sim \kyo \propto \gammaE\). The right vertical dotted line indicates the value of flow shear that is equal to the largest growth rate \(\gamma_\text{max}\), where the outer scale \(\kyo\) reaches, at least approximately, the scale of the most unstable mode \(k_{y, \text{max}}\rho_\perp \approx 3.7\). Note that, at low \(\gammaE\), \cref{eq:Q_poloidal_avg} is sometimes maximised at \(k_x = 0\). In those cases, represented by the hollow triangles, we have set \(\kxo\rho_\perp\) to the box scale \(2\pi \rho_\perp/L_x \approx 0.063\). }
	\label{fig:Q_vs_gammaE}
	\label{fig:outerscale_vs_gammaE}
\end{figure*}

Together, \cref{eq:phi}--\cref{eq:T} and \cref{eq:qn_etg} form an asymptotic model derived in an electrostatic, collisional limit of GK in a straight and uniform magnetic field with \(\uvec{b} = \hat{\vec{z}}\) \citep{adkins23}. They describe the electrostatic dynamics of fluctuations on perpendicular and parallel scales that satisfy
\begin{equation}
	\kpar L_T \sim \sqrt{\sigma}, \quad \kperp \rho_\perp \sim 1, \quad \rho_\perp \equiv \frac{\rhoe}{\sigma} \frac{L_T}{\lambda_{ei}},
\end{equation}
where \(L_T^{-1} \equiv -\partial \ln T_{e}/\partial x\) is the electron-temperature gradient, \(\lambda_{ei}\) is the electron-ion mean free path, and \(\sigma\) is a formal scaling parameter that is arbitrary provided it satisfies \(\beta_e \ll \sigma \ll 1\). The fact that this parameter is arbitrary is a consequence of the scale invariance of the model \citep{adkins23}. This implies a particular scaling of the heat flux with the square of the normalised parallel system size, viz., 
\begin{align}
	Q \propto \left(\frac{L_\parallel \sqrt{\sigma	}}{L_T}\right)^2.
	\label{eq:heatflux_lpar2}
\end{align}
Similarly, any intrinsic time scales in \cref{eq:phi}--\cref{eq:T} (e.g., the outer-scale injection rate \(\gammao\)) can be shown to be proportional to the inverse square of the normalised parallel box size. The numerical results for the (electron) heat flux \(Q\) and any relevant rate~\(\gamma\) (e.g., \(\gammaE, \gamma_\vk, \gammao\), etc) can, therefore, be presented in terms of the following normalised quantities:
\begin{align}
	\hat{Q} &\equiv \left(\frac{L_T}{L_\parallel \sqrt{\sigma}} \right)^2 \frac{Q}{(\rho_\perp/\rho_e) Q_{\text{gB}e}}, \label{eq:qnorm} \\
	\hat{\gamma} & \equiv \left(\frac{L_T}{L_\parallel \sqrt{\sigma}} \right)^{-2}  \frac{\gamma}{\omegaperp},
	\label{eq:gammanorm}
\end{align}
where \mbox{\(Q_{\text{gB}e} = n_{e} T_{e} \vthe (\rho_e/L_T)^2\)} is the gyro-Bohm heat flux and 
\begin{equation}
	\omegaperp = \rhoe \vthe / 2\rho_\perp L_T
	\label{eq:omegaperp_def}
\end{equation}
is the value of the electron drift frequency at $k_y \rho_\perp = 1$. A direct consequence of the scale invariance of \cref{eq:phi}--\cref{eq:T} is that \(\hat{Q}\) must be independent of \(L_T\) and the perpendicular box size (in any direction, provided \(L_x\) and \(L_y\) are sufficiently large); therefore, it is a function of \(\gammaEhat\) only. Note that all of the aforementioned scalings are valid only when the hyperviscous cutoff is far from the outer scale, i.e., when \(\nu_\perp\) is small enough and so does not upset the scale invariance of the outer-scale quantities.

In the absence of flow shear, the nonlinear saturated state of \cref{eq:phi}--\cref{eq:T} has been investigated extensively, and exhibits a critically balanced local energy cascade \citep{adkins23}. Therefore, \mbox{\cref{eq:phi}--\cref{eq:T}} with \(\gammaE\neq0\) should give rise to the kind of turbulence that is described by the theory laid out in \cref{sec:nonlinear_saturation}. Here, we show data from four sets of simulations where we varied \(\gammaE\) while keeping all other parameters fixed, as detailed in \cref{tab:sims}. 

\Cref{fig:Q_vs_gammaE}(a) shows the dependence of $\hat{Q}$ on \(\gammaEhat\) for these simulations. The agreement with \cref{fig:heatflux_anisotropic} is evident. The predicted scaling of the heat flux in the weak-shear regime \cref{eq:heatflux_weakly_sheared} holds very well up to \(\gammaEhat \approx 100\). This is followed by a swift transition to the strong-shear scaling \cref{eq:heatflux_strong_shear}. Recall that the theory of \cref{sec:weak_shear} predicts that the transition between the two regimes should occur at \(\gammaE \sim \gammaoO\), where the energy-injection rate at the outer scale \(\gammaoO\) is approximately the linear-instability growth rate \(\gamma_\vk\) at the outer scale. Using the outer-scale estimates presented in \cref{fig:outerscale_vs_gammaE}(b), we find that \(\kyoO\rho_\perp \approx 0.3\). Solving the linear dispersion relation for \cref{eq:phi}--\cref{eq:T} for Sim1 at \(k_x = 0\), \(k_y\rho_\perp=0.35\) and substituting \(\kpar L_T/\sqrt{\sigma} = 2\pi/50\) for the box-sized mode in the parallel direction, we obtain an approximation of the normalised outer-scale injection rate of \(\hat{\gamma}^\text{o}(0) \sim \hat{\gamma}_\vk \approx 70\), consistent with the numerically observed transition at \(\gammaEhat \approx 100\).\footnote{Given the discussion in \cref{footnote:gamma_ky}, we must mention that this \(\kpar\) maximises the growth rate at the given \(k_y\). The reader might also wonder why we consider \(k_x = 0\) given that the outer-scale radial wavenumber is not zero. Since \(k_x\) enters the linear dispersion relation only via the hyperviscous dissipation, the growth rate is virtually independent of \(k_x\) at these scales.}

\begin{figure}
	\centering
	\begingroup\import{figs/fig5}{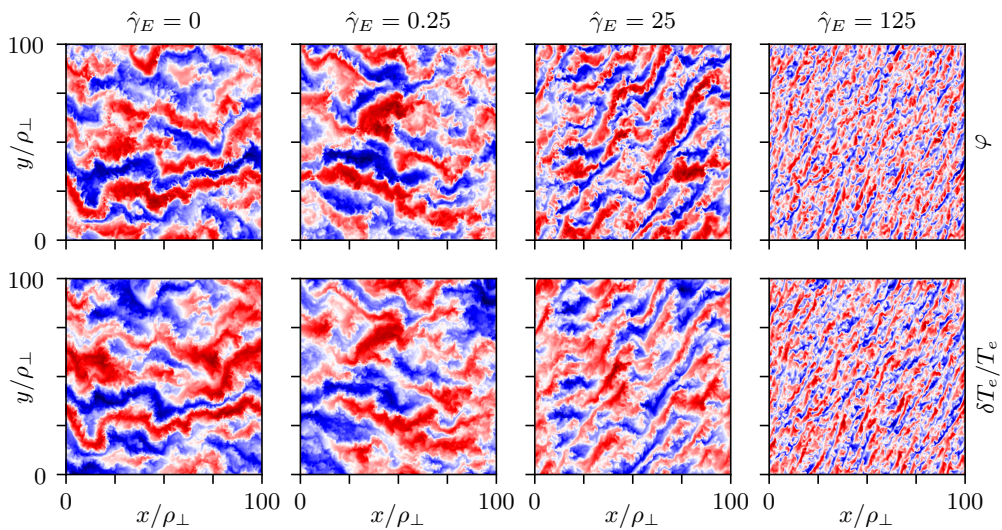}\endgroup
	\caption{Snapshots of \(\varphi\) (top row) and \(\delta T_e/T_{e}\) (bottom row) in the \((x, y)\) plane for Sim1 simulations with four different values of \(\gammaE\), as specified above each column. For each snapshot, the amplitudes are normalised to lie in the range \([-1, 1]\), with the values in this interval corresponding to colours between dark blue and dark red, respectively. The second column corresponds to the weak-shear regime (i) from \cref{fig:heatflux_anisotropic}(a), where the flow shear is too weak to influence the saturated state significantly. The third column also corresponds to the weak-shear regime, with \(\kyo(\gammaE)\) pinned to \(\kyoO\) but with \(\kxo(\gammaE)\) increased by the influence of the flow shear, which here clearly manifests itself as the tilting of the eddies. In this case, the structures have a similar size in $y$ to those in the first- and second-row panels, but a shorter length scale in $x$ due to being sheared. The last column shows the saturated state in the strong-shear regime~(ii) of \cref{fig:heatflux_anisotropic}(a), where the flow shear has manifestly pushed the outer scale to much shorter wavelengths.}
	\label{fig:tripleplot}
\end{figure}

\Cref{fig:outerscale_vs_gammaE}(b) shows that \(\kxo(\gammaE)\) and \(\kyo(\gammaE)\) are consistent with the behaviour of the outer scale predicted in \cref{sec:flow_shear}. As \(\gammaE\) is increased, \(\kyo\) stays roughly constant until \(\kxo\) catches up with it, whereafter the system transitions into the strong-shear regime where \(\kxo\) and \(\kyo\) remain comparable and both increase linearly with \(\gammaE\), as predicted in \cref{sec:strong_shear}. This behaviour persists until \(\kyo\) reaches (approximately) the poloidal wavenumber at which the linear growth rate is maximal. For \(\gammaE\) larger than \(\gamma_\text{max}\), the turbulence is completely suppressed. This behaviour is also visually confirmed by \cref{fig:tripleplot}, where we show real-space snapshots from simulations with different flow shear. 

The outer-scale wavenumbers shown in \cref{fig:outerscale_vs_gammaE} are defined as those that maximise the (steady-state) poloidally and radially averaged heat fluxes, defined respectively as
\begin{align}
	\langle Q\rangle_{y} (k_x) &\equiv \frac{3}{2} n_{e} T_{e} \vthe \sum_{k_y, \kpar} i k_y \rho_e{\phinorm}^*_\vk \frac{\delta{T}_{e, \vk}}{T_{e}}, \label{eq:Q_poloidal_avg} \\
	\langle Q\rangle_{x} (k_y) &\equiv \frac{3}{2} n_{e} T_{e} \vthe \sum_{k_x, \kpar} i k_y \rho_e{\phinorm}^*_\vk \frac{\delta{T}_{e, \vk}}{T_{e}} \label{eq:Q_radial_avg},
\end{align}
where \(\phinorm_\vk\) and \(\delta{T}_{e, \vk}\) are the three-dimensional Fourier amplitudes of the fields. The ratio of the heat flux at the transition to that at zero flow shear is found to be approximately \mbox{\(Q(0)/Q(\gammaEhat=100)\approx13\)}, which, according to \cref{eq:heatflux_weakly_sheared}, would require \(\AoO \approx 0.4\), consistent with \mbox{\(\kxo/\kyo \approx 0.3\)} as seen in \cref{fig:outerscale_vs_gammaE} at low values of flow shear. Given the remarkably good fit for \(Q\) as a function of \(\gammaE\), the small discrepancy is likely due to our estimates of \(\kxo\) and \(\kyo\) via \cref{eq:Q_poloidal_avg} and \cref{eq:Q_radial_avg} being an imperfect measure of the outer scale. For instance, \cref{eq:Q_poloidal_avg} fails to produce a nonzero estimate for \(k_x\) if \(\gammaE\) is too low or zero. Note that the scale invariance of the fluid system implies that the streamer aspect ratio \(\AoO\) of \cref{eq:phi}--\cref{eq:T} is not a function of any of the parameters of the system; i.e., it is just an order-unity (albeit measurably, and consequentially, smaller than unity) constant, so it is not possible to perform a scan in \(\AoO\).

\begin{figure}
	\centering
	\begingroup\import{figs/fig6}{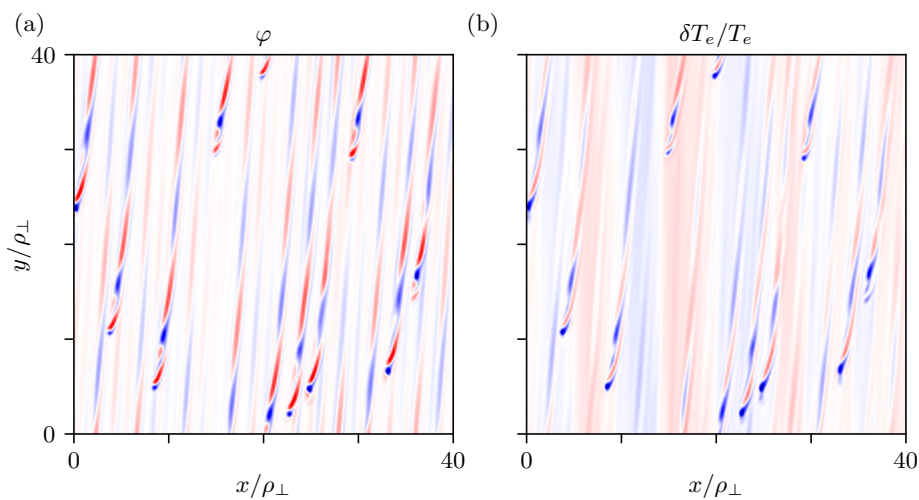}\endgroup
	\caption{Radial localisation of turbulent perturbations at very large values of flow shear. Taken from a Sim4 simulation with \(\gammaEhat = 540\), which is just over the largest growth rate \(\hat{\gamma}_\text{max} \approx 493\). The simulation has achieved a steady state with time-averaged normalised heat flux \(\hat{Q}(\gammaE)/\hat{Q}(0) \approx 4 \times 10^{-7}\), which is why it is not visible in \cref{fig:Q_vs_gammaE}.}
	\label{fig:etg_ferds}
\end{figure}

For values of \(\gammaE\) comparable to, or larger than, the maximum growth rate \(\gamma_\text{max}\), the turbulence is strongly suppressed, as expected. However, this occurs in a surprising and nontrivial way --- turbulence becomes radially localised into disjoint turbulent patches (see \cref{fig:etg_ferds}). This localisation is reminiscent of the formation of coherent structures called `ferdinons' in sheared ITG turbulence \citep{vanwyk2016, vanwyk2017, ivanov20}. Despite their qualitative similarities, the localised ETG structures reported here differ in a number of ways from the ITG ferdinons observed in similar fluid simulations by \citet{ivanov20}: the former are three-dimensional structures (i.e., require a \mbox{nonzero \(\kpar\))} that drift radially in the absence of magnetic drifts, while the latter exist in two dimensions and depend on the poloidal magnetic drift for their radial motion. The radial localisation of sheared turbulence through formation of coherent structures appears to be a universal phenomenon, whose detailed investigation is the subject of our ongoing work that falls outside the scope of this paper. Here, we note that, apart from the poloidally localised structures shown in \cref{fig:etg_ferds} and reported by \citet{vanwyk2016} and \citet{ivanov20}, solitary travelling structures that are localised only radially and consist of a single poloidal harmonic (i.e., a single \(k_y\)) have been seen in various models of sheared plasma turbulence \citep{mcmillan09,mcmillan18,pringle17,zhou19_zf,zhou20}. We find no such structures and it remains an open question exactly what the determining factor for their appearance is.

\subsection{Gyrokinetic ITG turbulence}
\label{sec:ITG}

We now explore the validity of our theory for plasma turbulence in axisymmetric toroidal geometry. Using the GK code \texttt{GENE}, we performed numerical flux-tube simulations of ITG-driven turbulence with modified adiabatic electrons \citep{hammett93} in a Cyclone-base-case (CBC) \citep{lin99,dimits00} geometry with normalised magnetic shear \(\hat{s}=0.796\), safety factor \(q = 1.4\), and inverse aspect ratio \(\epsilon = 0.18\). We focus on two values of the ion-temperature gradient, \mbox{\(R/L_{T_i} = 10\)} and \mbox{\(R/L_{T_i} = 14\)}, box sizes corresponding to the smallest wavenumbers \mbox{\(k_{x,\text{min}}\rhoi = 1.6\times10^{-2}\)} and \mbox{\(k_{y,\text{min}}\rhoi = 6.25\times10^{-3}\)}, \(n_x = 288\) radial modes, \(n_y=256\) poloidal modes for \(R/L_{T_i} = 10\) and \(n_y=512\) for \(R/L_{T_i} = 14\), \(n_z = 16\) parallel grid points, and velocity-space resolution \(n_v = 32\) (parallel velocity), \(n_\mu = 8\) (magnetic moment). We performed scans of the equilibrium perpendicular flow shear \(\gammaE\), with parallel flow shear turned off. Note that the temperature gradients that we consider are significantly above the nominal CBC value of \(R/L_{T_i} = 6.92\). This is done to ensure that the system is strongly driven and far from any marginal states where other physics, e.g., zonal flows, could be setting the saturated fluctuation levels.\footnote{Indeed, the assumption that saturation at the outer scale is determined by a balance between linear and nonlinear rates (see \cref{sec:energy_cascade}) is likely incorrect for ion-scale turbulence below the Dimits threshold \citep{dimits00}, where the fluctuations are regulated by strong long-lived zonal flows \citep{rogers00,diamond05,zhu20_prl,zhu20_jpp, ivanov20}. Investigating the influence of equilibrium flow shear on the Dimits state is outside of the scope of this paper.}

\begin{figure*}
	\centering
	\begingroup\import{figs/fig7}{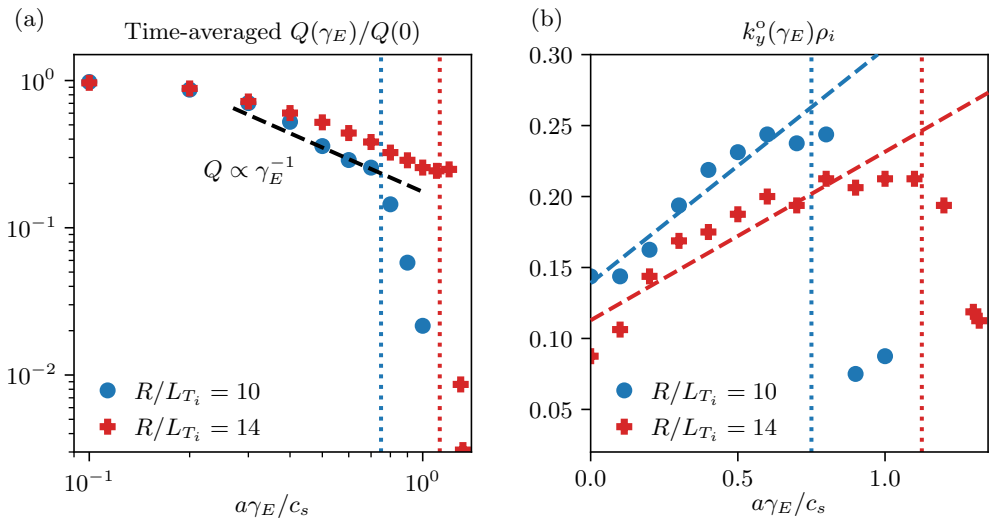}\endgroup
	\caption{(a) The time-averaged, saturated-state, radial turbulent heat flux, normalised to its value at \(\gammaE = 0\), and (b) the outer-scale poloidal wavenumber \(\kyo\rho_i\) as functions of the flow shear for two different values of the ion-temperature gradient (see \cref{sec:ITG} for other relevant numerical parameters). The black dashed line corresponds to the trend \(Q \propto \gammaE^{-1}\), while the blue and red dashed lines are linear fits for \(\kyo\) as a function of \(\gammaE\) for each temperature gradient. The vertical dotted lines correspond to \(1.5\gamma_\text{max}\) for each of the simulations. The flow shear is normalised to \(a/c_s\), where \(a\) is the minor radius and \(c_s\) is the ion sound speed.}
	\label{fig:GK_Q_vs_gammaE}
	\label{fig:GK_outerscale_vs_gammaE}
\end{figure*}

\Cref{fig:GK_Q_vs_gammaE} shows the dependence of the heat flux and poloidal outer-scale wavenumber on the shearing rate. The strong-shear scaling \(Q \propto \gammaE^{-1}\) \cref{eq:heatflux_strong_shear} is followed reasonably well for both values of the ion-temperature gradient. Also, the dependence of \(\kyo\) on \(\gammaE\) is approximately linear, as expected from~\cref{eq:kyo_strong_shear}. Note that \cref{eq:kyo_strong_shear} is only an asymptotic scaling, and for \(\gammaE\) close to \(\gammaoO\), we have \(\kyo(\gammaE) - \kyoO \propto \gammaE\). To calculate \(\kyo\) for \cref{fig:GK_outerscale_vs_gammaE}(b), we used the GK equivalent of \cref{eq:Q_radial_avg}, i.e., we  approximated \(\kyo\) by the \(k_y\) that has the largest contribution to the heat flux.

According to the theory presented in \cref{sec:strong_shear}, the strong-shear regime should end at \(\gammaE \sim \gamma_\text{max}\). Indeed, for both values of \(R/L_{T_i}\), the \(Q \propto \gammaE^{-1}\) dependence lasts roughly up to \(\gammaE/\gamma_\text{max} \approx 1.5\), after which the heat flux is sharply suppressed. Note the absence of the weak-shear scaling \cref{eq:heatflux_weakly_sheared} --- this is expected because the unsheared (\(\gammaE = 0\)) ITG turbulence in this case has \(\AoO \sim 1\). Consequently, the heat flux is (approximately) constant in the weak-shear regime and the dependence of \(Q\) on \(\gammaE\) resembles \cref{fig:heatflux_anisotropic}(b).

\subsubsection{Bistability of high-shear states}

\begin{figure}
	\begingroup\import{figs/fig8}{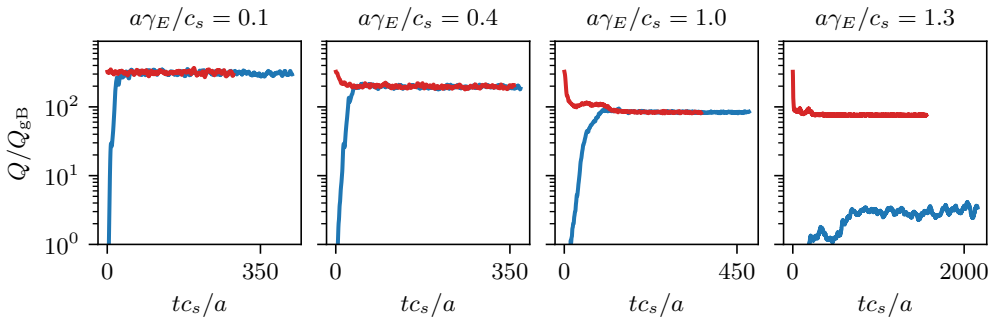}\endgroup
	\caption{Radial turbulent heat flux versus time for \mbox{\(R/L_{T_i} = 14\)} and four different values of \(\gammaE\), as labelled in the title of each panel. The blue lines are time traces from simulations initialised with small-amplitude noise, while the red ones represent simulations restarted from a saturated \(\gammaE=0\) run.}
	\label{fig:GK_hf}
\end{figure}

\begin{figure}
	\centering
	\begingroup\import{figs/fig9}{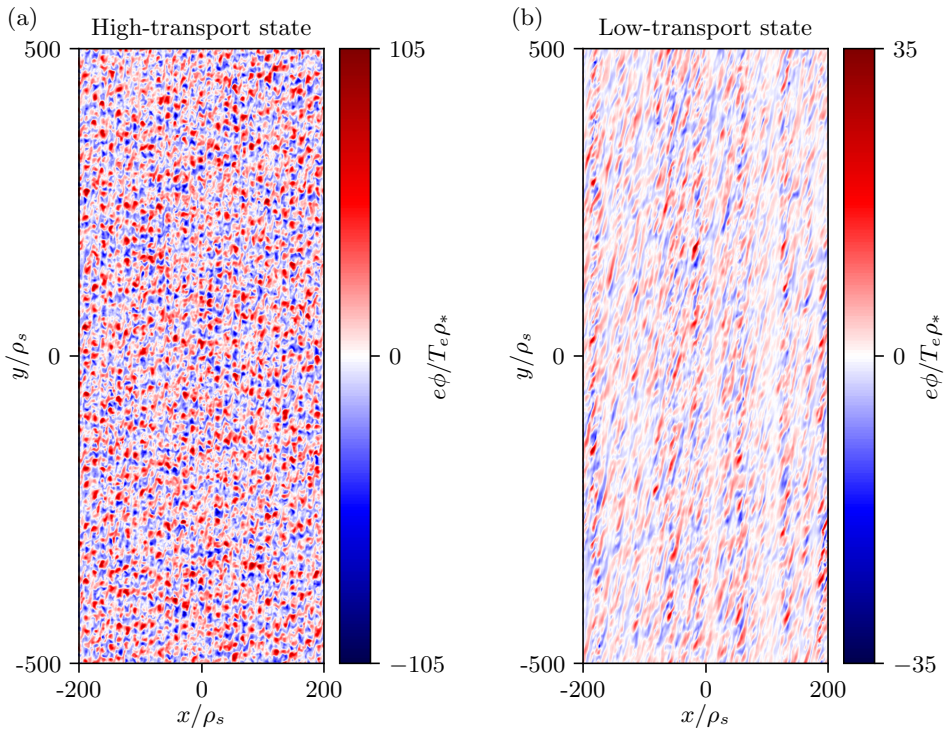}\endgroup
	
	\caption{Snapshots of the electrostatic potential in the perpendicular plane from the saturated high-transport (a) and low-transport (b) states with \(R/L_{T_i} = 14\) and \mbox{\(a\gammaE/c_s = 1.3\)}, whose heat-flux time traces are shown in blue and red, respectively, in the bottom right panel of \cref{fig:GK_hf}. The perpendicular coordinates are normalised to the ion sound radius \(\rho_s\). Note that the aspect ratio of the panels corresponds to that of the simulation domain. }
	\label{fig:GK_ferds}
\end{figure}

At large values of flow shear \(\gammaE > \gamma_\text{max}\), i.e., beyond the strong-shear regime, we find that the system can saturate at (at least) two different levels of heat transport. \Cref{fig:GK_hf} shows the time traces of the turbulent heat flux for \(R/L_{T_i} = 14\) and four different values of \(\gammaE\), where, for each value of the flow shear, the simulations were initialised either with small-amplitude noise or with data from a saturated \(\gammaE = 0\) simulation. For \(a\gammaE/c_s \leq 1\), the saturated state is found to be independent of the initial conditions. In contrast, a simulation with \(a\gammaE/c_s = 1.3\) initialised with a small-amplitude noise (corresponding to the rightmost point in \cref{fig:GK_Q_vs_gammaE}) saturates with a time-averaged radial heat flux that is nearly two orders of magnitude smaller than one obtained by restarting it from an already saturated high-amplitude state. Similar bistability in gyrokinetic turbulence with mean flow shear has been reported by \citet{christen22}. The physics of this phenomenon, as observed in the simulations presented in \cref{fig:GK_hf}, falls outside of the range of validity of the theory presented in \cref{sec:nonlinear_saturation} because, at least in the case investigated here, it happens only at \(\gammaE > \gamma_\text{max}\), where the assumption of a balance between the rates of shearing and energy injection \cref{eq:isotropic_outer_scale_balance} is not expected to hold.

Finally, we note that the appearance of coherent structures (whether of the poloidally localised kind or not, see the discussion at the end of \cref{sec:ETG}) can naturally lead to nonunique saturation. Indeed, if the saturated state is a collection of localised structures, then volume-averaged turbulent quantities, like the heat flux, are proportional to the number of structures in the simulation domain \citep{vanwyk2016}. If these structures happen to be well localised and non-interacting (or only weakly interacting), then their number in the simulation domain is not necessarily uniquely determined, but can be a function of the initial conditions and/or of the perpendicular box size. The turbulent fluctuations in the low-transport saturated state displayed in \cref{fig:GK_ferds} show signs of spatial localisation (as opposed to those in the high-transport state), even if not quite as obviously as they do in the ETG turbulence shown in \cref{fig:etg_ferds}.

\section{Momentum transport}
\label{sec:momflux}

We have thus far focused on the influence of imposed flow shear on the turbulent heat transport. In sheared systems, another important quantity of interest is the turbulent transport of momentum, which is crucial for driving and maintaining equilibrium differential rotation in fusion experiments. Such rotation is typically achieved using neutral beams that deposit both energy and momentum into the plasma. The profiles of temperature and equilibrium flow are then determined by the turbulent heat diffusivity and turbulent viscosity, which are usually larger than their collisional counterparts. The simplifying choice of a purely perpendicular flow made in \cref{sec:flow_shear} means that we are unable to describe all of the relevant physics: e.g., we are missing the effect of the parallel-velocity-gradient instability, which can be the main driver of transport at high \(\gammaE\) \citep{highcock10, barnes_shear11} (see also \cref{footnote:pvg}). Nevertheless, the theory derived in \cref{sec:flow_shear} makes numerically falsifiable predictions for the momentum transport of turbulence with imposed flow shear. Let us investigate them here.

Consider the fluid ETG model described in \cref{sec:ETG}, wherein we have imposed a mean flow in the poloidal \(y\) direction that varies along the radial \(x\) direction. This flow then allows the plasma to have nonzero radial transport of poloidal momentum, defined, similarly to \cref{eq:heatflux_def}, as
\begin{equation}
	\Pi \equiv \sum_\s m_\s \int \frac{\rmd^3 \vec{r}}{V} \intw (\vec{v}_E \bcdot \grad x) (\vec{w} \bcdot \grad y) \dfs. \label{eq:momflux_def}
\end{equation}
It can be shown (see \cref{appendix:momflux_ETG}) that, in the fluid model described in \cref{sec:ETG}, \cref{eq:momflux_def} becomes
\begin{equation}
	\Pi = -\frac{n_{e} T_{e} \rhoe^2 }{2}  \int \frac{\rmd^3 \vec{r}}{V} \frac{\partial \phinorm}{\partial y} \frac{\partial}{\partial x} \left[\left(1 + \frac{Z T_{e}}{T_{i}}\right) \phinorm - \frac{\delta T_e}{T_{e}}\right].
	\label{eq:momflux_etg}
\end{equation}
As expected, \(\Pi\) is written as the sum of the Reynolds stress of the \(E\times B\) flow and a diamagnetic stress \citep{smolyakov00, ivanov20, sarazin21}. Note that \(\Pi\) itself is of no relevance to the dynamics of the fluctuations as it does not enter \cref{eq:phi}--\cref{eq:T} at all, unlike \(Q\), which is responsible for the injection of energy, as per \cref{eq:free_energy}.

Just like we did for the heat flux \(Q\) in \cref{sec:heatflux}, we can estimate \(\Pi\) as
\begin{equation}
	\Pi \sim n_{e} T_{e} \kxotilt \kyo \rhoe^2 (\phinormko)^2,
	\label{eq:momflux_estimate}
\end{equation}
where
\begin{equation}
	\kxotilt(\gammaE) \equiv \kxo(\gammaE) - \kxoO \sim \kyo(\gammaE)\taunlo(\gammaE)\gammaE.
	\label{eq:kxotilt}
\end{equation}
Using \(\kxotilt\) instead of \(\kxo\) in \cref{eq:momflux_estimate} is necessary because, in the absence of flow shear, the GK equation \cref{eq:gk} obeys the symmetry
\begin{equation}
	(x, y, z) \mapsto (-x, y, -z), \ w_\parallel \mapsto -w_\parallel, \ \phi \mapsto -\phi, \ h_\s \mapsto -h_\s,
	\label{eq:updown_symmetry}
\end{equation}
under which \(\Pi \mapsto -\Pi\) and so the time-averaged \(\Pi\) must vanish if \(\gammaE = 0\) \citep{parra11, sugama11, fox17}. Thus, only the part of \(k_x\) associated with the eddy tilting contributes to the momentum flux.\footnote{A careful reader may spot another issue, which is also resolved by using \(k_{x,\text{tilt}}\) instead of \(k_x\). Recall the discussion after \cref{eq:heatflux_estimate}, where we justified estimating \(Q\) via its value at the outer scale by arguing that the contributions from smaller scales decay with increasing \(k_y\). If we were to estimate {na\"ively} the contribution \(\overline{\Pi}(k_y)\) to \(\Pi\) from scale \(k_y\) in the inertial range as \(\overline{\Pi}(k_y) \propto k_x k_y \phinormk^2\), the inertial-range spectrum \cref{eq:inertial_range_scaling} would imply \(\overline{\Pi}(k_y) \propto k_y^{2/3}\), leading one to believe that the integral \cref{eq:momflux_def} is dominated by the small-scale end of the inertial range rather than by the outer scale. This argument is, however, incorrect. In the inertial range, as \(k_y\) increases, the nonlinear time decreases according to \cref{eq:inertial_range_scaling}, limiting the effect of the flow shear and thus decreasing \(\overline{\Pi}(k_y)\). To capture this, the correct estimate is \(\overline{\Pi}(k_y) \propto k_{x, \text{tilt}} k_y \phinormk^2\), where \(k_{x, \text{tilt}} \sim k_y \taunl \gammaE \propto k_y^{-1/3}\) in the inertial range. This then leads to \(\overline{\Pi}(k_y) \propto k_y^{-2/3}\), and so \(\Pi\) is dominated by the contributions are the outer scale.}

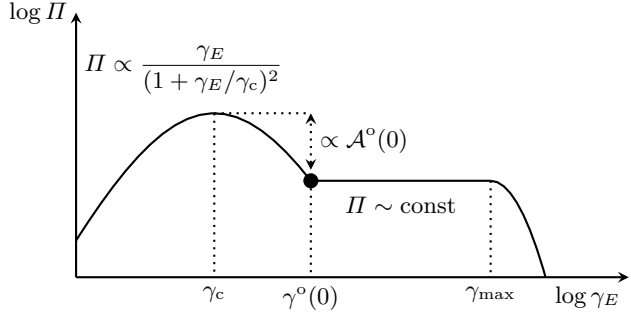
\begin{figure}
	\centering
	\scalebox{1.0}{\begin{tikzpicture}[scale=1, thick, every node/.style={scale=1}]
    \begin{loglogaxis}[width=3.5in, height=2in, clip mode=individual,
    axis lines=left,
    axis line style={thick},
    xmin=0.01,xmax=100,
    xlabel=$\log \gammaE$,ylabel=$\log \Pi$,
    ymin=0.03,ymax=0.3,
    ticklabel style={font=\tiny,fill=white},
    xtick={\empty},ytick={\empty},
    xlabel style={at={(ticklabel* cs:1)},anchor=north east},
    ylabel style={at={(ticklabel* cs:1)},anchor=east,rotate=-90}
    ]
    \def\xc{0.1}
    \def\xo{0.5}
    \def\xendstrong{10}
    \def\xendflat{80}
    \def\Ampl{5}
    \addplot[samples=20,domain=0.01:\xo,black,thick] {\Ampl*x/(1 + x/\xc)^2};
    \addplot[samples=20,domain=\xo:\xendstrong,black,thick] {\Ampl*\xo/(1 + \xo/\xc)^2};
    \addplot[samples=20,domain=\xendstrong:\xendflat,black,thick] {\Ampl*\xo*exp( -(ln(\xendstrong/x))^2)/(1 + \xo/\xc)^2};
    
    \node[circle,inner sep=2pt,fill=black] at (\xo,{\Ampl*\xo/(1+\xo/\xc)^2}) {};
    
    \draw[-, dotted, thick] (\xc,\Ampl*\xc/4) --node[pos=1,anchor=north]{\(\gammaani\)} (\xc,0.03);
    \draw[-, dotted, thick] (\xo,{\Ampl*\xo/(1+\xo/\xc)^2}) --node[pos=1,anchor=north]{\(\gammaoO\)} (\xo,0.03);
    \draw[-, dotted, thick] (\xendstrong,{\Ampl*\xo/(1+\xo/\xc)^2}) --node[pos=1,anchor=north]{\(\gamma_\text{max}\)} (\xendstrong,0.03);
    \draw[-, dotted, thick] (\xc,\Ampl*\xc/4) -- (\xo,\Ampl*\xc/4);
    
    \node[anchor=south] at (0.6*\xc, 1.1*\Ampl*\xc/4) {
    	$\begin{aligned}
    		\Pi \propto \frac{\gammaE}{(1+\gammaE/\gammaani)^2}
    	\end{aligned}$
    	};
    
    \node[anchor=south] at ({sqrt(\xo*\xendstrong)}, {0.7*\Ampl*\xo/(1 + \xo/\xc)^2}) {\(\Pi \sim \text{const}\)};
    
    \draw[<->, >=stealth, dotted, thick] (\xo,{1.1*\Ampl*\xo/(1+\xo/\xc)^2}) -- node[midway, right, scale=1] {
    	$\begin{aligned}
    		\propto \AoO
    	\end{aligned}$
    } (\xo,\Ampl*\xc/4);
\end{loglogaxis}

\end{tikzpicture}}
	\caption{A qualitative diagram of the momentum flux \(\Pi\) vs. flow shear \(\gammaE\) in the fluid ETG model (see \cref{fig:Q_vs_gammaE} for a similar diagram for the heat flux \(Q\)). The indicated ratio between the plateau in the strong-shear regime and the peak of \(\Pi\) at \(\gammaE=\gammaani\) formally holds when \(\AoO \ll 1\). }
	\label{fig:momflux}
\end{figure}

Using \cref{eq:taunl_A} and \cref{eq:outer_scale_def}, we write \cref{eq:momflux_estimate} as
\begin{equation}
	\frac{\Pi}{n_{e} T_{e}} \sim \frac{\gammaE \gammao}{\Omega_e^2} \frac{1}{(\kxo \rhoe)^2}\sim \frac{\gammaE \gammao}{\Omega_e^2} \frac{1}{(\Ao \kyo \rhoe)^2}.
	\label{eq:momflux_estimate_nicer}
\end{equation}
In the weak-shear regime (\cref{sec:weak_shear}), where \mbox{\(\gammaE < \gammaoO\)}, the linear and nonlinear times at the outer scale are given by \cref{eq:outer_scale_balance_weak_shear}, and the fluctuation aspect ratio at the outer scale satisfies \cref{eq:kxo_weak_shear}, we find 
\begin{equation}
	\frac{\Pi(\gammaE)}{\Pi(0)} \propto \frac{\gammaE}{(1 + \gammaE/\gammaani)^2}.
	\label{eq:momflux_weakshear}
\end{equation}
Note that if \(\AoO\sim1\), i.e., if the turbulence is isotropic in the absence of flow shear, then \(\gammaani\sim\gammaoO\) and \(\Pi\) is (approximately) proportional to \(\gammaE\) in the weak-shear regime, unlike \(Q\), which would be constant in this case [see \cref{fig:heatflux_anisotropic}(b)]. In the strong-shear regime (\cref{sec:strong_shear}), where \mbox{\(\gammaE > \gammaoO\)}, the fluctuations satisfy \(\Ao(\gammaE) \sim 1\), and \cref{eq:isotropic_outer_scale_balance} and \cref{eq:kyo_strong_shear} hold, we find
\begin{equation}
	\Pi(\gammaE) \propto \gammaE^0,
	\label{eq:momflux_strongshear}
\end{equation}
i.e., the momentum flux is independent of the imposed flow shear (see \cref{fig:momflux}). Crucially, in both regimes,
\begin{equation}
	\frac{\Pi}{Q} \propto \gammaE.
	\label{eq:pi_over_q}
\end{equation}
This can also be seen directly from \cref{eq:heatflux_estimate_fancy} and \cref{eq:momflux_estimate_nicer}, which imply
\begin{equation}
	\frac{\Pi \vthe}{Q} \sim \frac{\gammaE}{\gammao} \kyo \rhoe,
\end{equation}
and so \cref{eq:pi_over_q} follows immediately if \(\gammao \propto \kyo\), as we have assumed throughout \cref{sec:flow_shear}. If we define the normalised turbulent viscosity \(\nu\) and diffusivity \(\chi\) as
\begin{equation}
	\nu \equiv \frac{\Pi}{n_{e} m_e \gammaE}, \quad
	\chi \equiv \frac{Q}{n_{e} T_{e}L_T^{-1}},
\end{equation}
then the turbulent Prandtl number is
\begin{equation}
	\text{Pr} \ = \frac{\nu}{\chi} = \frac{\Pi}{Q} \frac{\vthe^2}{2\gammaE L_T}.
	\label{eq:pr_nice}
	\end{equation}
According to \cref{eq:pi_over_q}, this is independent of \(\gammaE\). This prediction is consistent with previous GK studies of turbulent transport in sheared turbulence \citep{highcock10, barnes_shear11}.

\begin{figure}
	\centering
	\begingroup\import{figs/fig11}{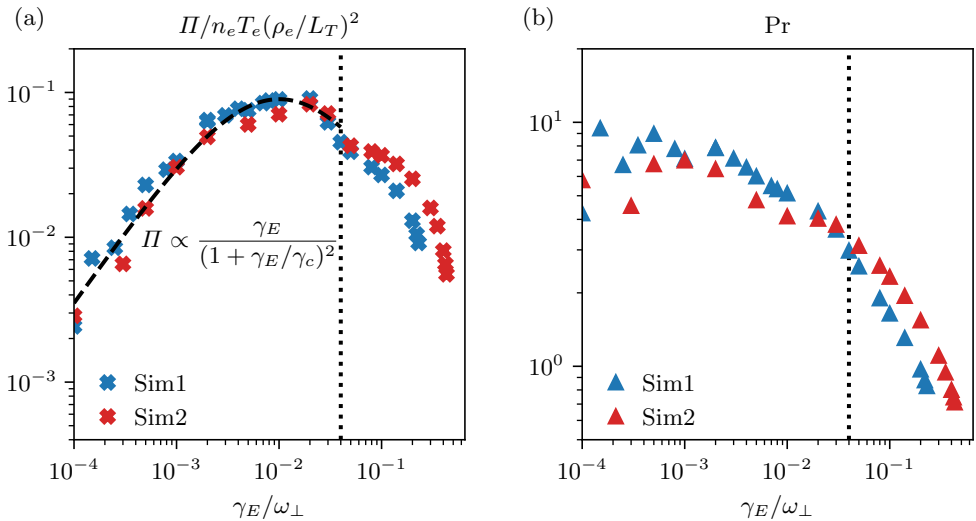}\endgroup
	\caption{(a) Radial flux of poloidal momentum in the fluid ETG model \cref{eq:momflux_etg} as a function of \(\gammaE\) from simulation sets Sim1 and Sim2 (see \cref{tab:sims}). The black dashed line is the best-fit line of the form \cref{eq:momflux_weakshear} to the data up to \(\gammaE/\omega_\perp = 0.04\) (denoted by the vertical dotted line), where the system transitions from the weak- to the strong-shear regime (see also \cref{fig:Q_vs_gammaE}). (b) Prandtl number, defined as \cref{eq:pr_nice} for the same simulations as in panel (a).}
	\label{fig:momflux_pr}
\end{figure}

\Cref{fig:momflux_pr}(a) shows the values of turbulent heat and momentum flux from the Sim1 and Sim2 sets of simulations, detailed in \cref{tab:sims}. There is good agreement with the weak-shear scaling \cref{eq:momflux_weakshear}. According to \cref{eq:momflux_weakshear}, \(\Pi\) should peak in the weak-shear regime at \(\gammaE = \gammaani = \AoO\gammaoO\). For the Sim1 and Sim2 simulations, whose data is presented in \cref{fig:momflux_pr}, the transition to the strong-shear regime happens at \(\gammaoO/\omega_\perp \approx 0.04\), where \(\omega_\perp\) is given by \cref{eq:omegaperp_def} (see also \cref{fig:Q_vs_gammaE}). As discussed in \cref{sec:ETG}, we find that, numerically, \(\AoO \approx 0.4\), which gives \(\gammaani/\omega_\perp \approx 0.016 \), consistent with the observed peak in \(\Pi\) at \(\gammaE/\omega_\perp \approx 0.01\). However, the prediction that \(\Pi\) should be constant in the strong-shear regime is not observed, likely due to finite-hyperviscosity effects and the finite extent of the inertial range. Our theory does not take into account the finite value of hyperviscosity necessary to achieve a saturated state numerically. Since \(\Pi\) is a higher-order Fourier-space moment of the fluctuation spectrum than \(Q\), it is more sensitive to the spectrum at high \(k\), where hyperviscous effects matter. These effects become more important as \(\gammaE\) increases and the inertial range shortens (see also \cref{fig:flowshear_spectra} in \cref{appendix:aniso_to_iso}). As \cref{fig:momflux_pr}(a) shows, the simulations with lower hyperviscosity (Sim2) have a weaker dependence of \(\Pi\) on \(\gammaE\) in the strong-shear regime, consistent with the hypothesis that the hyperviscous cutoff is responsible for the discrepancy with the theoretical prediction \cref{eq:momflux_strongshear}. \Cref{fig:momflux_pr}(b) shows that the numerically measured Prandtl number is weakly dependent on \(\gammaE\), varying only by about a factor of two over a range for \(\gammaE\) of nearly three orders of magnitude, before decreasing at large \(\gammaE\) where hyperviscous effects become important.

The nonmonotonic weak-shear dependence \cref{eq:momflux_weakshear}, which has a local maximum at \mbox{\(\gammaE = \gammaani\)}, implies that, for the same value of \(\Pi\), two distinct values of \(\gammaE\), and so of \(Q\), are possible. This suggests that sheared anisotropic turbulence may be prone to transport bifurcations similar to those discussed by \citet{highcock10, highcock11} and \citet{parra11_shear}. Of course, our oversimplified model of electron-scale turbulence cannot be applied directly to any experimental studies. Nevertheless, our theory suggests that if experimentally relevant turbulence is dominated by streamers, transport bifurcations might exist.

\section{Summary and discussion}
\label{sec:summary}

Starting from the standard picture of turbulent saturation via a local energy cascade (\cref{sec:energy_cascade}), we have developed a theory for the effect of mean perpendicular flow shear on temperature-gradient-driven turbulence in fusion plasmas. As argued in \cref{sec:flow_shear}, it is meaningful to distinguish two different regimes depending on the ratio of the shearing rate \(\gammaE\) to the rate \(\gammaoO\) of energy injection in the corresponding system with zero flow shear. 

In the weak-shear regime, defined by \(\gammaE < \gammaoO\), the poloidal outer scale and the energy-injection rate remain approximately the same as when \(\gammaE = 0\), but the radial outer scale decreases with increasing \(\gammaE\) \cref{eq:kxo_weak_shear}, due to the tilting of turbulent eddies by the shear. The extent to which the flow shear is able to suppress the turbulent transport in this regime is linked to the aspect ratio of the outer-scale fluctuations at zero flow shear, \(\AoO = \kxoO/\kyoO\). We find that turbulence with \(\AoO \sim 1\) is largely unaffected by flow shear unless the shear is comparable to, or larger than, \(\gammaoO\). In contrast,  heat transport in streamer-dominated turbulence with \(\AoO \ll 1\), which is often encountered in fusion-relevant contexts, is shown to be strongly suppressed by flow shear even at \(\gammaE \ll \gammaoO\) as long as \(\gammaE \gtrsim \gammaani\), where the critical shearing rate \(\gammaani\) \cref{eq:gammaani_def} is smaller than \(\gammaoO\) by a factor of \(\AoO\). This reflects the intuitive notion that radially elongated fluctuations should be more susceptible to sheared poloidal flows. 

At \(\gammaE > \gammaoO\), the system is in the strong-shear regime, where the outer scale is determined by the balance between the shearing, energy-injection, and nonlinear mixing rates \cref{eq:isotropic_outer_scale_balance}, and the outer-scale perpendicular wavenumbers grows proportionally to \(\gammaE\). Turbulence in the strong-shear regime is found always to have \(\kxo \sim \kyo\), even if it was dominated by streamers at \(\gammaE=0\). This is due to the shear-induced tilting of the turbulent fluctuations, which forces them to have \(\Ao(\gammaE) \sim 1\) in this regime.

Our theoretical predictions for the dependence of the radial turbulent heat flux on the rate of perpendicular flow shear, \cref{eq:heatflux_weakly_sheared} and \cref{eq:heatflux_strong_shear}, are confirmed to hold over a range of four orders of magnitude for the flow shear in idealised fluid ETG simulations (\cref{sec:ETG}). Additionally, GK flux-tube simulations of ITG turbulence have demonstrated that our theory applies to more realistic models of plasma turbulence as well (\cref{sec:ITG}). These two models are paradigmatic cases of turbulence with \(\AoO \ll 1\) and \(\AoO \sim 1\), respectively.

In addition to the heat-flux scalings, in \cref{sec:momflux}, we utilise our theory to predict the dependence of momentum flux on flow shear in the fluid ETG simulations. While not directly applicable to more realistic numerical simulations due to the restriction of purely perpendicular flow shear, our results suggest that streamer-dominated turbulent transport might exhibit transport bifurcations.

There exists a body of work on the suppression of turbulence by flow shear that is based on `decorrelation theories' \citep{shaing90, biglari90, zhang92, zhang93, hatch18}  stemming from \citet{dupree72} or the equivalent derivation by \citet{zhang93}. There are some parallels that can be drawn between these theories and our approach: e.g., the definition of the normalised flow shear \(W\) by \citet{zhang92} and \citet{hatch18} includes the crucial influence of the fluctuation aspect ratio and can be mapped to \(\gammaE / \Ao(\gammaE) \gammaoO\) in our notation. However, the derivation of our theory, as presented in \cref{sec:flow_shear}, is not related to these decorrelation theories and produces different scalings for the dependence of the heat flux on \(\gammaE\). For example, the functional dependence~\cref{eq:heatflux_weakly_sheared}, borne out by the numerical results presented in \cref{sec:ETG}, is not reproduced by these theories. While the decorrelation theory by \citet{zhang92} has been claimed to be applicable to some GK simulations \citep{hatch18}, we are (at this stage) unable to draw any conclusions about the more general applicability either of decorrelation theories or of the theory presented in \cref{sec:flow_shear}.

One possible application of the theory of sheared streamer-dominated turbulence described in \cref{sec:weak_shear} is in the description of multiscale interactions in magnetised-plasma turbulence. Gyrokinetic theory and numerical analysis point to the existence of linear instabilities and turbulent fluctuations at two disparate perpendicular scales, viz., the gyroradii of the main ion species \(\rhoi\) and of the electrons \(\rhoe\), often referred to as the `ion' and `electron' (perpendicular) scales, respectively. By virtue of the difference in the ion and electron masses, these two scales are well separated, viz., \(\rhoe/\rhoi \sim \sqrt{m_e/m_i} \ll 1\). Furthermore, the GK ordering implies that the growth rate of electron-scale instabilities satisfies \(\gamma_e \sim \vthe/L\), while that of the ion-scale instabilities is \mbox{\(\gamma_i \sim \vthi/L\)}, where \(L\) is some appropriate measure of the size of the device. Therefore, the fluctuations at these two spatial scales also occur on well-separated time scales: \(\gamma_i/\gamma_e \sim \sqrt{m_i / m_e} \ll 1\). Nevertheless, numerically expensive GK simulations with enough resolution to span both ion and electron scales have shown that ion-scale turbulence can suppress electron-scale fluctuations \citep{candy07, waltz07, maeyama15, howard16a}, and experimental results in support of this hypothesis have been reported \citep{howard16b}. 

One possible mechanism for this suppression is that the turbulent ion-scale \exb{} flows shear away the electron-scale perturbations. However, if we assume that the electron-scale fluctuations satisfy \(\A_{e} \sim 1\), this could not happen: the time scale associated with the ion-scale \exb{} shear is \(\gamma_i\), while the growth rate of the electron-scale instabilities is \(\gamma_e \gg \gamma_i\), so, by the quench rule, \(\gamma_i\) is too small to suppress the electron-scale turbulence.\footnote{Note that there are other mechanisms that could be responsible for such suppression, e.g., the parallel variation of the ion-scale \exb{} flow \citep{hardman19, hardman20}, or ITG turbulence giving rise to shearing flows of small enough radial scale to influence the ETG eddies \citep{holland04}.}

The theory of weakly sheared turbulence proposed in \cref{sec:weak_shear} offers a possible alternative explanation. Simulations of electron-scale turbulence indicate that turbulence at those scales is dominated by radially elongated streamers, and, according to our theory, the ion-scale \exb{} shear will be relevant for shearing these streamers if they have \(\A_{e} \ll 1\). In particular, if \(k_{y, e} \rhoe \sim 1\) but \(k_{x, e} \rhoi \sim 1\), i.e., if the ETG streamers have a radial size comparable to the scale of ion fluctuations, then the critical shear rate \(\gamma_{\text{c},e}\) for ETG turbulence \cref{eq:gammaani_def} would be of the order of the ion-scale fluctuating \exb{} shear. Such an ordering would put the ion-scale shearing rate in the weak-shear regime for the ETG streamers, where \cref{eq:heatflux_weakly_sheared} holds, with \(\gammaE\) now representing the value of the flow shear that electron-scale fluctuations experience from the ion-scale ones. Therefore, the influence of this flow shear would be non-negligible, despite the na\"ive argument that might lead one to presume otherwise. Numerical experiments have indeed shown that the inclusion of ion-scale zonal flows (flux-surface-constant, poloidal \exb{} shear flows) in ETG simulations can drastically reduce the levels of electron-scale turbulence and transport, even though the ion-scale zonal-flow shear is much smaller than the growth rate of electron-scale instabilities \citep{waltz07}. Whether this mechanism for ITG suppression of electron-scale fluctuations is indeed realised in multiscale plasma turbulence appears to be a promising subject for future work.

\section*{Acknowledgements}
We thank M. R. Hardman and F. I. Parra for inspiring discussions and invaluable feedback. We would also like to thank the organisers of the 14\textsuperscript{th} Plasma Kinetics Working Meeting (2023) and its hosts at the Wolfgang Pauli Institute in Vienna, where a significant part of this work was first presented and discussed.

\section*{Funding}	
This work was supported by the Engineering and Physical Sciences Research Council (EPSRC) [EP/R034737/1 and EP/W006839/1]. PGI and AAS were also supported in part by the Simons Foundation via a Simons Investigator award to AAS. TA acknowledges the support of the Royal Society Te Ap\=arangi, through Marsden-Fund grant MFP-UOO2221. Simulations were performed using resources provided by the Cambridge Service for Data Driven Discovery (CSD3) operated by the University of Cambridge Research Computing Service (\href{https://www.csd3.cam.ac.uk}{www.csd3.cam.ac.uk}), provided by Dell EMC and Intel using Tier-2 funding from the Engineering and Physical Sciences Research Council (capital grant EP/T022159/1), and DiRAC funding from the Science and Technology Facilities Council (\href{https://www.dirac.ac.uk}{www.dirac.ac.uk}). To obtain further information on the data and models underlying this paper contact \href{mailto:PublicationsManager@ukaea.uk}{PublicationsManager@ukaea.uk}.

\section*{Declaration of interests}
The authors report no conflict of interest.

\appendix

\section{Isotropisation of streamer-dominated turbulence in the inertial range}
\label{appendix:aniso_to_iso}

Here we develop the theory of the transition range between a streamer-dominated outer scale with \mbox{\(\Ao \ll 1\)} and an inertial range with \(\A \sim 1\), in the weak-shear regime, as promised in \cref{sec:weak_shear}. To be more specific, the transition range is a poloidal-scale range \mbox{\(\kyo(\gammaE) \lesssim k_y \lesssim k_y^\text{T}\)}, wherein the fluctuations' aspect ratio increases from \(\A(\kyo) \ll 1\) at the outer scale to \(\A(k_y^\text{T}) \sim 1\) at the end of the transition range, located at a new scale \mbox{\(k_y \sim k_y^\text{T}\)}. Note that, by definition, free-energy injection by the linear instability is negligible at all scales below the outer scale.

First, let us show that radial wavenumbers below the outer scale cannot be determined by the tilting of the eddies by the flow shear, i.e., that \(k_x\) must be determined by nonlinear effects (with a boundary condition \(k_x = \kxo\) at the outer scale). Consider the simpler case of \(\AoO \ll 1\) turbulence in the presence of a flow shear of intermediate strength \(\gammaani \ll \gammaE \ll \gammaoO\). In this case, the radial wavenumber at the outer scale is given by~\cref{eq:kx_weak_shear}:
\begin{equation}
	\kxo(\gammaE) \sim \kyoO \taunloO \gammaE.
	\label{eq:appendix_kx_outer_intermediate_shear}
\end{equation}
Now suppose that \cref{eq:appendix_kx_outer_intermediate_shear} holds beyond the outer scale, viz., that
\begin{equation}
	k_x \sim k_y \taunl \gammaE
	\label{eq:appendix_kx_intermediate_shear}
\end{equation}
for \(k_y > \kyo\). Combining \cref{eq:appendix_kx_intermediate_shear} with the definition of the nonlinear time \cref{eq:taunl} and assuming that the free-energy flux \cref{eq:constant_energy_cascade} is constant below the outer scale, we find that, for the scales where \cref{eq:appendix_kx_intermediate_shear} holds,
\begin{equation}
	\taunl^{-1} \propto \gammaE^{2/5} k_y^{4/5} \propto k_y^{4/5}.
	\label{eq:taunl_too_slow_appendix}
\end{equation}
However, beyond the outer scale, the nonlinear mixing rate must increase faster with \(k_y\) than the injection rate \(\gamma_\vk \propto k_y\), which \cref{eq:taunl_too_slow_appendix} does not. Therefore, the assumption that \cref{eq:appendix_kx_intermediate_shear} holds beyond the outer scale contradicts the very definition of the outer scale that we adopted in \cref{sec:energy_cascade}.

Instead, let us suppose that, due to nonlinear mixing beyond the outer scale, the radial wavenumbers corresponding to the poloidal wavenumbers in the range \mbox{\(\kyo(\gammaE) \lesssim k_y \lesssim k_y^\text{T}\)} satisfy
\begin{equation}
	\frac{k_x}{\kxo} \sim \left(\frac{k_y}{\kyo}\right)^{1+\lambda}.
	\label{eq:kx_sigma}
\end{equation}
The parameter \(\lambda\) is a measure of the tendency of the nonlinear mixing to `isotropise' (or `anisotropise' if \mbox{\(\lambda < 0\)}) the turbulent fluctuations with increasing \(k_y\).\footnote{Notice that, as discussed in \cref{sec:weak_shear}, one possibility is that the outer-scale aspect ratio \(\AoO\) is inherited by the inertial-range cascade in the sense that \(\A\) in the inertial range is independent of \(k_y\), i.e., it is scale invariant. In the analysis above, this corresponds to \(\lambda = 0\). In this case, the transition range is, in fact, the entire inertial range, \(\A\) is scale-invariant by \cref{eq:A_sigma}, and the spectra \cref{eq:transition_spectrum_ky} and \cref{eq:transition_spectrum_kx} agree with \cref{eq:inertial_range_scaling}.} This can also be expressed as
\begin{equation}
	\frac{\A(k_y)}{\Ao} \sim \left(\frac{k_y}{\kyo}\right)^\lambda.
	\label{eq:A_sigma}
\end{equation} 
If \(\lambda > 0\), the transition region ends at \(\A(k_y^\text{T}) \sim 1\), so \cref{eq:A_sigma} gives us
\begin{equation}
	k_y^\text{T}  \sim \frac{\kyo}{(\Ao)^{1/\lambda}} \gg \kyo.
\end{equation}

Inside the transition region, the expression for the nonlinear time \cref{eq:taunl_A} and the assumption of constant free-energy flux \cref{eq:constant_energy_cascade} jointly imply
\begin{equation}
	\frac{\phinormk}{\phinormko} \sim \left(\frac{k_y}{\kyo}\right)^{-2/3 \ - \ \lambda/3}.
	\label{eq:transition_spectrum_ky}
\end{equation}
Therefore, the \(k_y\) spectrum in the transition range must be steeper than the inertial-range spectrum \cref{eq:inertial_range_scaling}. Using \cref{eq:kx_sigma}, we can recast \cref{eq:transition_spectrum_ky} in terms of the radial wavenumbers:
\begin{equation}
	\frac{\phinormk}{\phinormko} \sim \left(\frac{k_x}{\kxo}\right)^{-2/3 \ + \ \lambda/3(1+\lambda)},
	\label{eq:transition_spectrum_kx}
\end{equation}
which is shallower than the inertial-range spectrum \cref{eq:inertial_range_scaling}. 

\begin{figure}
	\centering
	\begingroup\import{figs/fig12}{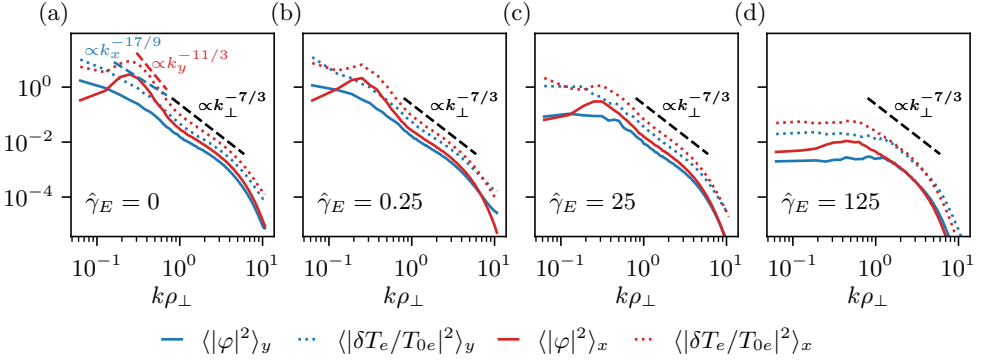}\endgroup
	\caption{Panels (a)--(d) shows the spectra \cref{eq:phi_vs_ky_def}--\cref{eq:T_vs_kx_def}, as indicated in the legend at the bottom, for the four simulations shown in \cref{fig:tripleplot}. All panels show an inertial-range spectrum that agrees with the predicted \(\propto \kperp^{-7/3}\) scaling, shown as a black dashed line. In panel (a), where \(\gammaE = 0\) and the transition range is widest, we also show the predicted transition-range scalings in the case \(\lambda = 2\) of the spectrum with \(k_x\) and \(k_y\) in blue and red dashed lines, respectively, with the exponents labelled accordingly.}
	\label{fig:flowshear_spectra}
\end{figure}

We do not know how to determine \(\lambda\) theoretically but we can confirm numerically that the above arguments are sound and that \(\lambda > 0\). In \cref{fig:flowshear_spectra}, we show the spectra of \(\phinorm\) and \(\delta T_e/T_{e}\) from the ETG fluid model discussed in \cref{sec:ETG} for the four Sim1 simulations with varying flow shear presented in \cref{fig:tripleplot}. As a proxy for the dependence of \(\phinormk\) and \(\overline{\delta T_e}/T_{e}\) on \(k_x\) and \(k_y\), we are using the following averages:
\begin{align}
	\langle|\phinorm|^2\rangle_x(k_y) &\equiv \sum_{k_x, \kpar} |\phinorm_\vk|^2, \label{eq:phi_vs_ky_def} \\
	\langle|\phinorm|^2\rangle_y(k_x) &\equiv \sum_{k_y, \kpar} |\phinorm_\vk|^2, \label{eq:phi_vs_kx_def}  \\
	\left\langle\left\vert\frac{\delta T_e}{T_{e}}\right\vert^2\right\rangle_x(k_y) &\equiv \sum_{k_x, \kpar} \left\vert\frac{\delta T_{e,\vk}}{T_{e}}\right\vert^2, \label{eq:T_vs_ky_def} \\
	\left\langle\left\vert\frac{\delta T_e}{T_{e}}\right\vert^2\right\rangle_y(k_x) &\equiv \sum_{k_y, \kpar} \left\vert\frac{\delta T_{e,\vk}}{T_{e}}\right\vert^2. \label{eq:T_vs_kx_def}
\end{align}
Note that \cref{eq:phi_vs_ky_def}--\cref{eq:T_vs_kx_def} differ from \cref{eq:phinormk_def} by a factor of \(k_x\) or \(k_y\). Thus, the expected scalings that follow from \cref{eq:transition_spectrum_ky} and \cref{eq:transition_spectrum_kx} are 
\begin{align}
	&\langle|\phinorm|^2\rangle_x \sim \left\langle\left\vert\frac{\delta T_e}{T_{e}}\right\vert^2\right\rangle_x \propto k_y^{-7/3  \ - \ 2\lambda/3}, \\
	&\langle|\phinorm|^2\rangle_y \sim \left\langle\left\vert\frac{\delta T_e}{T_{e}}\right\vert^2\right\rangle_y \propto k_x^{-7/3 \ + \ 2\lambda/3(1+\lambda)}.
\end{align}
\Cref{fig:flowshear_spectra}(a) shows that the fluid-ETG spectra obtained from solving \cref{eq:phi}--\cref{eq:T} with \(\gammaE = 0\) are roughly consistent with \(\lambda = 2\). Increasing the value of the flow shear reduces the aspect ratio, as expected, and flattens the \(k_y\) spectrum to approximately match the \(k_x\) one. Unfortunately, because the aspect ratio \(\AoO\) cannot be varied in the fluid model, we are unable to vary the size of the transition range and its small span in wavenumbers limits our ability to measure \(\lambda\) numerically and to verify the transition-range theory laid out above with any greater accuracy. Nevertheless, our numerical results are consistent with it.

\section{Numerical implementation of the fluid model}
\label{appendix:fluid_numerics}

The numerical results presented in \cref{sec:ETG} are obtained by solving \cref{eq:phi}--\cref{eq:T} in the shearing frame
\begin{equation}
	t' = t, \ x' = x, \ y' = y - x\gammaE t, \ z' = z,
	\label{eq:appendix_shearing_frame}
\end{equation}
instead of the laboratory frame \((x, y, z)\). By using a pseudo-spectral algorithm and solving for the evolution of Fourier modes in \((x', y', z')\), we impose triply periodic boundary conditions in the shearing frame. The linear terms in the equations are integrated using an implicit Crank-Nicolson method, while the nonlinear ones are integrated explicitly using the Adams-Bashforth three-step method. This code is a modification of that developed and used by \citet{ivanov20}, \citet{ivanov22} and \citet{adkins23}.

Our approach for dealing with the time-dependent radial derivatives arising from \cref{eq:appendix_shearing_frame} differs from the usual spectral remapping scheme that was first proposed by \mbox{\citet{hammett_flowshear_06}} for the GK code \texttt{GS2} and that forms the basis of the implementation of equilibrium flow shear in many modern gyrokinetics codes, e.g., \texttt{stella} \citep{barnes_stella19, stonge22}, \texttt{GKW} \citep{peeters_gkw09}, and \texttt{GENE} \citep{told_thesis12}, the latter being used for the numerical simulations presented in \cref{sec:ITG}. While this remapping algorithm is fairly robust, the approximations that it makes can, in some cases, lead to unphysical results \citep{mcmillan19}. Even though steps can be taken to improve the algorithm \citep{mcmillan19, christen21}, the simplicity of \mbox{\cref{eq:phi}--\cref{eq:T}} allows us to take a different, both simpler and more robust, approach. We consider a fixed \((k_x', k_y')\) grid of shearing-frame wavenumbers, which corresponds to continuously drifting laboratory-frame radial wavenumbers \mbox{\(k_x = k_x' - \gammaE k_y' t\)} that are periodic (in time) on the interval  \mbox{\(-K_x \leq k_x \leq K_x\)}, where \mbox{\(K_x \equiv \pi N_x/L_x\), \(L_x\)} is the radial box size, and \(N_x\) is the `padded' number of radial wavenumbers. By ensuring that \mbox{\(N_x \geq \lfloor3n_x/2\rfloor\)}, where \(n_x\) is the number of resolved radial wavenumbers, we eliminate aliasing issues in the pseudo-spectral method by zeroing out all modes with \(|k_x| > k_{x, \text{max}} \equiv \pi n_x / L_x\) --- this is the standard `\(2/3\) rule' \citep{orszag71}. For example, the simulations in the Sim2 set (see \cref{tab:sims}) have \(n_x = 683\) and \(N_x = 1024\).

\begin{figure}
	\centering
	\hspace{-2mm}\scalebox{1.0}{\begin{tikzpicture}[scale=1, thick, every node/.style={scale=1}]
	\def\N{16}
	\def\halfN{8}
	\def\n{10}
	\def\halfn{5}
	\def\L{20}
	\def\Kmax{2*pi*\halfN/\L}
	\def\kmax{2*pi*\halfn/\L}
	\def\xlim{1.2*\Kmax}
	\def\ylim{1.5}
	\def\kone{-0.55}
	\def\ktwo{-0.27}
	\def\ksum{\kone+\ktwo}
	
	\begin{axis}[width=3.7in, height=1in, clip mode=individual,
		axis lines=center,
		axis line style={thick},
		xmin=-\xlim,xmax=\xlim,
		xlabel=$k_x$,
		ymin=0,ymax=\ylim,
		xtick={-\Kmax, -\kmax, \kmax, \Kmax, \kone, \ktwo, \ksum}, ytick={\empty},
		xticklabels={$-K_{x}$, $-k_{x, \text{max}}$, $k_{x, \text{max}}$, $K_{x}$, $k_1$, $k_2$, $k_3$},
		xlabel style={at={(ticklabel* cs:1)},anchor=north west}
		]
		
		\node at (-\xlim, \ylim) {(a)};
		
		\addplot[samples=200,domain=-\xlim:\xlim,black,thick] {exp(-(x+0.2)^2-x^4)};
		
		\draw[-, solid, thick] (-\Kmax, 0) -- (-\Kmax, \ylim);
		\draw[-, solid, thick] (\Kmax, 0) -- (\Kmax, \ylim);
		
		\draw[-, dashed, thick] (-\kmax, 0) -- (-\kmax, \ylim);
		\draw[-, dashed, thick] (\kmax, 0) -- (\kmax, \ylim);
		
		\draw[-, dotted, thick] (\kone, 0) -- (\kone, \ylim);
		\draw[-, dotted, thick] (\ktwo, 0) -- (\ktwo, \ylim);
		\draw[-, dotted, thick] (\ksum, 0) -- (\ksum, \ylim);
		
		\def\yoffset{0.1}
		\draw[-stealth, solid] (\kone, \ylim-\yoffset) to [out=210,in=-30] (\ksum, \ylim-\yoffset);
		\draw[-stealth, solid] (\ktwo, \ylim) to [out=150,in=30] (\ksum, \ylim);
		
		\fill [gray, opacity=0.5] (-\Kmax, 0) rectangle (-\kmax,\ylim);
		\fill [gray, opacity=0.5] (\kmax, 0) rectangle (\Kmax,\ylim);
	\end{axis}
\end{tikzpicture}
	}
	\hspace{-2mm}\scalebox{1.0}{\begin{tikzpicture}[scale=1, thick, every node/.style={scale=1}]
	\def\N{16}
	\def\halfN{8}
	\def\n{10}
	\def\halfn{5}
	\def\L{20}
	\def\Period{2*pi*\N/\L}
	\def\period{2*pi*\n/\L}
	\def\Kmax{2*pi*\halfN/\L}
	\def\kmax{2*pi*\halfn/\L}
	\def\xlim{1.2*\Kmax}
	\def\ylim{1.5}
	
	\begin{axis}[width=3.7in, height=1in, clip mode=individual,
		axis lines=center,
		axis line style={thick},
		xmin=-\xlim,xmax=\xlim,
		xlabel=$k_x'$,
		ymin=0,ymax=\ylim,
		xtick={-\Kmax, -\kmax-\dx+\Period, \kmax-\dx, \Kmax}, ytick={\empty},
		xticklabels={$-K_{x}$, $-k_{x, \text{max}}$, $k_{x, \text{max}}$, $$},
		xlabel style={at={(ticklabel* cs:1)},anchor=north west}
		]
		
		\node at (-\xlim, \ylim) {(b)};
		
		\def\dx{1}
		\addplot[samples=200,domain=-\Kmax:\Kmax-\dx,black,thick] {exp(-(x+0.2+\dx)^2-(x+\dx)^4)};
		\addplot[samples=200,domain=\Kmax-\dx:\Kmax,black,thick] {exp(-(x+0.2+\dx-\Period)^2-(x+\dx-\Period)^4)};
		
		\draw[-, solid, thick] (-\Kmax, 0) -- (-\Kmax, \ylim);
		\draw[-, solid, thick] (\Kmax, 0) -- (\Kmax, \ylim);
		
		\draw[-, dashed, thick] (-\kmax-\dx+\Period, 0) -- (-\kmax-\dx+\Period, \ylim);
		\draw[-, dashed, thick] (\kmax-\dx, 0) -- (\kmax-\dx, \ylim);
		
		\fill [gray, opacity=0.5] (-\Kmax, 0) rectangle (-\kmax,\ylim);
		\fill [gray, opacity=0.5] (\kmax, 0) rectangle (\Kmax,\ylim);
	\end{axis}
\end{tikzpicture}}
	\hspace{-2mm}\scalebox{1.0}{\begin{tikzpicture}[scale=1, thick, every node/.style={scale=1}]
	\def\N{16}
	\def\halfN{8}
	\def\n{10}
	\def\halfn{5}
	\def\L{20}
	\def\Period{2*pi*\N/\L}
	\def\period{2*pi*\n/\L}
	\def\Kmax{2*pi*\halfN/\L}
	\def\kmax{2*pi*\halfn/\L}
	\def\xlim{1.2*\Kmax}
	\def\ylim{1.5}
	\def\kone{-0.55}
	\def\ktwo{-0.25}
	\def\ksum{\kone+\ktwo}
	
	\def\dx{1}
	
	\def\axisshift{-1cm}
	
	\begin{axis}[width=3.7in, height=1in, clip mode=individual, name=topaxis,
		axis lines=center,
		axis line style={thick},
		xmin=-\xlim,xmax=\xlim,
		xlabel=$k_x'$,
		ymin=0,ymax=\ylim,
		xtick={\kone-\dx, \ktwo-\dx,\ksum-2*\dx+\period}, ytick={\empty},
		xticklabels={$k_1$, $k_2$, ${\color{red} k_3}$},
		xlabel style={at={(ticklabel* cs:1)},anchor=north west}
		]
		
		\def\axislabeloffset{0}
		\node at (-\xlim, \ylim+\axislabeloffset) {(c)};
		
		
		\addplot[samples=200,domain=-\kmax:\kmax-\dx,black,thick] {exp(-(x+0.2+\dx)^2-(x+\dx)^4)};
		\addplot[samples=200,domain=\kmax-\dx:\kmax,black,thick] {exp(-(x+0.2+\dx-\period)^2-(x+\dx-\period)^4)};
		
		\draw[-, solid, thick] (-\Kmax, 0) -- (-\Kmax, \ylim);
		\draw[-, solid, thick] (\Kmax, 0) -- (\Kmax, \ylim);
		
		\draw[-, dashed, thick] (\kmax-\dx, 0) -- (\kmax-\dx, \ylim);
		
		\draw[-, dotted, thick] (\kone-\dx, 0) -- (\kone-\dx, \ylim);
		\draw[-, dotted, thick] (\ktwo-\dx, 0) -- (\ktwo-\dx, \ylim);
		\draw[-, dotted, thick] (\ksum-2*\dx, 0) -- (\ksum-2*\dx, \ylim);
		\draw[-, dotted, thick] (\ksum-2*\dx+\Period, 0) -- (\ksum-2*\dx+\Period, \ylim);
		
		\def\yoffset{0.1}
		\draw[-stealth, solid] (\kone-\dx, \ylim-\yoffset) to [out=210,in=-30] (\ksum-2*\dx, \ylim-\yoffset);
		\draw[-stealth, solid] (\ktwo-\dx, \ylim) to [out=150,in=30] (\ksum-2*\dx, \ylim);
		\draw[-stealth, solid] (\ksum-2*\dx, \ylim) to [out=10,in=170] (\ksum-2*\dx+\Period, \ylim);
		
		\addplot[samples=200,domain=-\kmax:\kmax-2*\dx,red,thick] {exp(-(x+0.2+2*\dx)^2-(x+2*\dx)^4)};
		\addplot[samples=200,domain=\kmax-2*\dx:\kmax,red,thick] {exp(-(x+0.2+2*\dx-\period)^2-(x+2*\dx-\period)^4)};
		\draw[-, dashed, thick, red] (\kmax-2*\dx, 0) -- (\kmax-2*\dx, \ylim);
		\draw[-, dotted, thick, red] (\ksum-2*\dx+\period, 0) -- (\ksum-2*\dx+\period, \ylim);

		\fill [gray, opacity=0.5] (-\Kmax, 0) rectangle (-\kmax,\ylim);
		\fill [gray, opacity=0.5] (\kmax, 0) rectangle (\Kmax,\ylim);
	\end{axis}
	
\end{tikzpicture}}
	\hspace{-2mm}\scalebox{1.0}{\begin{tikzpicture}[scale=1, thick, every node/.style={scale=1}]
	\def\N{16}
	\def\halfN{8}
	\def\n{10}
	\def\halfn{5}
	\def\L{20}
	\def\Period{2*pi*\N/\L}
	\def\period{2*pi*\n/\L}
	\def\Kmax{2*pi*\halfN/\L}
	\def\kmax{2*pi*\halfn/\L}
	\def\xlim{1.2*\Kmax}
	\def\ylim{1.5}
	
	\def\kone{-0.55}
	\def\ktwo{-0.25}
	\def\ksum{\kone+\ktwo}
	\def\dx{1}
	\def\axisshift{-1cm}
	
	\begin{axis}[width=3.7in, height=1in, clip mode=individual,
		axis lines=center,
		axis line style={thick},
		xmin=-\xlim,xmax=\xlim,
		xlabel=$k_x'$,
		ymin=0,ymax=\ylim,
		xtick={\kone-\dx, \ktwo-\dx,\ksum-2*\dx+\Period}, ytick={\empty},
		xticklabels={$k_1$, $k_2$, ${\color{red}k_3}$},
		xlabel style={at={(ticklabel* cs:1)},anchor=north west},
		name=topaxis
		]
		
		\def\axislabeloffset{0}
		\node at (-\xlim, \ylim+\axislabeloffset) {(d)};
		
		
		\addplot[samples=200,domain=-\Kmax:\Kmax-\dx,black,thick] {exp(-(x+0.2+\dx)^2-(x+\dx)^4)};
		\addplot[samples=200,domain=\Kmax-\dx:\Kmax,black,thick] {exp(-(x+0.2+\dx-\Period)^2-(x+\dx-\Period)^4)};
		
		\draw[-, solid, thick] (-\Kmax, 0) -- (-\Kmax, \ylim);
		\draw[-, solid, thick] (\Kmax, 0) -- (\Kmax, \ylim);
		
		\draw[-, dashed, thick] (-\kmax-\dx+\Period, 0) -- (-\kmax-\dx+\Period, \ylim);
		\draw[-, dashed, thick] (\kmax-\dx, 0) -- (\kmax-\dx, \ylim);
		
		\draw[-, dotted, thick] (\kone-\dx, 0) -- (\kone-\dx, \ylim);
		\draw[-, dotted, thick] (\ktwo-\dx, 0) -- (\ktwo-\dx, \ylim);
		
		\def\yoffset{0.1}
		\draw[-stealth, solid] (\kone-\dx, \ylim) to [out=210,in=-30] (\ksum-2*\dx, \ylim-\yoffset);
		\draw[-stealth, solid] (\ktwo-\dx, \ylim) to [out=150,in=30] (\ksum-2*\dx, \ylim);
		
		\draw[-stealth, solid] (\ksum-2*\dx, \ylim) to [out=10,in=170] (\ksum-2*\dx+\Period, \ylim);
		
		\fill [gray, opacity=0.5] (\kmax-\dx, 0) rectangle (-\kmax-\dx+\Period,\ylim);
		
		
		\addplot[samples=200,domain=-\Kmax:\Kmax-2*\dx,red,thick] {exp(-(x+0.2+2*\dx)^2-(x+2*\dx)^4)};
		\addplot[samples=200,domain=\Kmax-2*\dx:\Kmax,red,thick] {exp(-(x+0.2+2*\dx-\Period)^2-(x+2*\dx-\Period)^4)};
		
		\draw[-, dashed, thick, red] (-\kmax-2*\dx+\Period, 0) -- (-\kmax-2*\dx+\Period, \ylim);
		\draw[-, dashed, thick, red] (\kmax-2*\dx, 0) -- (\kmax-2*\dx, \ylim);
		
		\draw[-, dotted, thick] (\ksum-2*\dx, 0) -- (\ksum-2*\dx, \ylim);
		\draw[-, dotted, thick, red] (\ksum-2*\dx+\Period, 0) -- (\ksum-2*\dx+\Period, \ylim);
		
		\fill [red, opacity=0.3] (-\Kmax-2*\dx+\Period, 0) rectangle (-\kmax-2*\dx+\Period,\ylim);
		\fill [red, opacity=0.3] (\kmax-2*\dx, 0) rectangle (\Kmax-2*\dx,\ylim);
	\end{axis}

\end{tikzpicture}}
	\caption{(a) Typical fluctuation amplitudes of sheared turbulence with \(k_y=q\) for some fixed \(q\) (note that the spectrum does not peak at \(k_x\) = 0 if \(\gammaE \neq 0\)) as a function of the laboratory-frame radial wavenumber \(k_x\). The shaded regions indicate the zero padding needed for dealiasing. The vertical solid and vertical dashed lines denote \(\pm K_x\) and \(\pm k_{x, \text{max}}\), respectively. Two radial wavenumbers, \(k_1\) and \(k_2\), are shown, together with \(k_3 = k_1 + k_2\) into which they couple nonlinearly. (b)~The same fluctuation amplitudes as in (a), but now transformed to the shearing frame using \cref{eq:kx_drift_appendix}, while keeping the padded region fixed to the outer \(1/3\) of the wavenumbers in the shearing frame. (c) Same as (b) but with \(k_x\) made periodic on \([-k_{x, \text{max}}, k_{x, \text{max}}]\) and computed via \cref{eq:kx_drift_appendix_smallerperiod} instead. The fluctuation amplitudes with \(k_y = 2q\) are shown in red. Here, \(k_1\) and \(k_2\) are the same two modes as in panel (a) that couple nonlinearly into \(k_3\) (shown in red as the corresponding poloidal wavenumber is \(k_y=2q\)). However, in this version of dealiasing, their sum falls into the dealiased region. (d) The correct way to represent the fluctuations in the shearing frame using \cref{eq:kx_drift_appendix} and zeroing out modes with \(|k_x| > k_{x, \text{max}}\). We are showing the same two wavenumbers \(k_1\) and \(k_2\) as before. They couple nonlinearly to modes with \((k_x, k_y) = (k_3, 2q)\), whose fluctuation amplitudes and dealiased regions are shown in red. In (c) and (d), to illustrate the correspondence between modes with \(k_y = q\) and \(k_y = 2q\), we have used the exact same function of \(k_x\) to represent fluctuation amplitudes at both poloidal wavenumbers.}
	\label{fig:aliasing}
\end{figure}
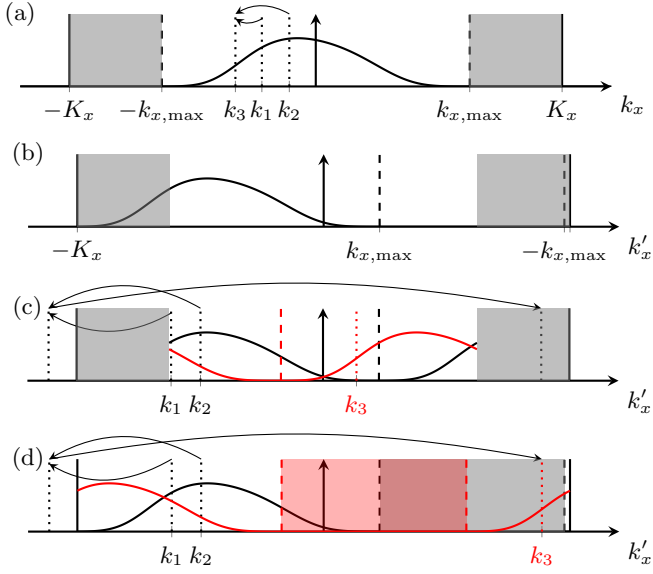

This scheme leads to a minor complication that is not encountered if one uses the remapping algorithm instead. To understand this, consider that, even in the presence of flow shear, the fluctuation amplitudes peak near \(k_x = 0\) and are suppressed at large \(|k_x|\) because the dissipation terms (whether physical or numerical) are determined by the laboratory-frame wavenumber \(k_x\), not by the shearing-frame one \(k_x'\). Thus, the appropriate choice for a Galerkin truncation, required by any pseudo-spectral algorithm, should be based on the laboratory-frame wavenumbers. This laboratory-frame truncation, along with the necessary zero padding for dealiasing, is illustrated in \cref{fig:aliasing}(a) in the case of zero flow shear. However, for \(\gammaE \neq 0\), we wish to solve the equations in terms of \(k_x'\) rather than \(k_x\). To do this, we assign a laboratory-frame~\(k_x\) to each \(k_x'\) using
\begin{equation}
	k_x = k_x' - \gammaE k_y' t + \frac{2\pi N_x m}{L_x},
	\label{eq:kx_drift_appendix}
\end{equation}
where \(m\) is an integer chosen so that \(-K_x \leq k_x \leq K_x\). \Cref{fig:aliasing}(b) shows that using \cref{eq:kx_drift_appendix} na\"ively can (depending on how much \(k_x\) has drifted) bring physically meaningful small-\(|k_x|\) modes into the region where the amplitudes are zeroed out to avoid aliasing. Effectively, this na\"ive method offsets the interval of resolved laboratory-frame modes so that, for any given \(k_y \neq 0\), it is no longer centred around \(k_x = 0\). Therefore, this approach does not work. One alternative is to consider instead \(k_x\) that drifts periodically as
\begin{equation}
	k_x = k_x' - \gammaE k_y' t + \frac{2\pi n_x m}{L_x},
	\label{eq:kx_drift_appendix_smallerperiod}
\end{equation}
where now \(m\) is chosen so that \mbox{\(-k_{x, \text{max}} \leq k_x \leq k_{x, \text{max}}\)}, i.e., the radial wavenumbers drift periodically only within the physically resolved interval \([-k_{x, \text{max}}, k_{x, \text{max}}]\), instead of the larger `padded' interval \([-K_x, K_x]\). However, this method is not suitable for calculating the nonlinear terms: as \cref{fig:aliasing}(c) shows, it removes physically meaningful nonlinear couplings. Similarly, one can also show that it also introduces aliasing-like couplings between physically unrelated modes (e.g., one can choose modes with \(k_1, \ k_2 > 0\) that are nonlinearly coupled to \mbox{\(k_1 + k_2 - 2\pi n_x/L_x < 0\)}). The correct approach, which employs \cref{eq:kx_drift_appendix} and dealiasing based on \(|k_x|\), is demonstrated in \cref{fig:aliasing}(d). This requires keeping track of the laboratory-frame \(k_x\) and zeroing out modes based on it. Fortunately, this is straightforward to do, and the additional overhead of having to compute the inverse of the linear response, required for the implicit Crank-Nicolson method, is negligible for the simple fluid model. 

Note that continuously drifting wavenumbers have one extra advantage over the remapping algorithm. A mode that has been sheared so much that it reemerges at the other side of the resolved wavenumber interval has necessarily passed through the dealiased region. Thus, its amplitude is zero and so cannot introduce any spurious fluctuations.

This algorithm is effectively the same as that used by \citet{lithwick07_shear} (and attributed by him to Gordon Ogilvie) for simulations of hydrodynamic sheared flows.

\section{Momentum flux in the fluid ETG model}
\label{appendix:momflux_ETG}

We want to compute the radial flux of poloidal momentum,
\begin{equation}
	\Pi \equiv \sum_\s m_\s \int \frac{\rmd^3 \vec{r}}{V} \intw (\vec{v}_E \bcdot \grad x) (\vec{w} \bcdot \grad y) \dfs,
	\label{eq:momflux_def_appendix}
\end{equation}
in the fluid ETG model used in \cref{sec:ETG} and \cref{sec:momflux}. The details of the derivation of the model can be found in Appendix B of \citet{adkins23}. Here we require only a few facts about the limit in which this derivation was carried out. 

First, the ion gyroradius is large, viz., \(\kperp \rhoi \gg 1\), which allows us to neglect \(h_i\) in the ion distribution function \(\dfi\) given by \cref{eq:delta_f}, so
\begin{equation}
	\dfi \approx -\frac{Ze \phi}{T_{i}} \Fi.
\end{equation}
This implies that the ion contribution to \cref{eq:momflux_def_appendix} vanishes:
\begin{equation}
	\int \frac{\rmd^3 \vec{r}}{V} \intw (\vec{v}_E \bcdot \grad x) (\vec{w} \bcdot \grad y) \dfi \propto \int \frac{\rmd^3 \vec{r}}{V} \frac{\partial \phi}{\partial y} \phi = 0.
\end{equation}
Secondly, the electron distribution function is expressed as 
\begin{equation}
	\dfe = \phinorm\Fe + h_e,
	\label{eq:dfe_form}
\end{equation}
where \(\phinorm = e \phi / T_{e}\) and \(h_e\) is gyroangle independent at fixed \(\vec{R}_e = \vec{r} - \hat{\vec{z}}\times\vec{w}/\Omega_e\). Since the radial \(\vec{E}\times \vec{B}\) velocity is
\begin{equation}
	\vec{v}_E \bcdot \grad x = -\frac{1}{2}\rhoe\vthe\frac{\partial \phinorm}{\partial y},
\end{equation}
\cref{eq:momflux_def_appendix} becomes
\begin{equation}
	\Pi = -\frac{1}{2}\rhoe m_e\vthe \int \frac{\rmd^3 \vec{r}}{V} \intw \frac{\partial \phinorm}{\partial y}(\vec{w} \bcdot \grad y) h_e,
	\label{eq:momflux_derivation_firststep}
\end{equation}
where the contribution from the \(\phinorm\Fe\) part of \cref{eq:dfe_form} has vanished. Evaluating \cref{eq:momflux_derivation_firststep} is a standard GK calculation, which we outline now. 

Noting that the perpendicular part of \(\vec{w}\) is
\begin{equation}
	\vec{w}_\perp = w_\perp (\cos\theta \hat{\vec{x}} - \sin\theta \hat{\vec{y}}),
\end{equation}
where \(\theta\) is the gyroangle, and that
\begin{equation}
 	\hat{\vec{z}} \times \vec{w}_\perp = -\frac{\partial \vec{w}_\perp}{\partial \theta},
\end{equation}
we find
\begin{align}
	&\intw \vec{w}_\perp h_e\!= \! \intw \hat{\vec{z}}\times \frac{\partial \vec{w}_\perp}{\partial \theta} h_e\!=\!-\!\intw \hat{\vec{z}}\times\vec{w}_\perp \frac{\partial h_e}{\partial \theta}.
	\label{eq:momflux_derivation_thirdstep}
\end{align}
As \(h_e\) is independent of \(\theta\) at fixed \(\vec{R}_e\), the last partial derivative in \cref{eq:momflux_derivation_thirdstep} (which is evaluated at fixed \(\vec{r}\)) becomes
\begin{equation}
	\frac{\partial h_e}{\partial \theta} = \frac{\partial}{\partial\theta}\left(-\frac{\hat{\vec{z}}\times\vec{w}}{\Omega_e}\right) \bcdot \frac{\partial h_e}{\partial \vec{R}_e} = - \frac{\vec{w}_\perp}{\Omega_e}\bcdot \frac{\partial h_e}{\partial \vec{R}_e}.
	\label{eq:he_theta_derivative}
\end{equation}
Substituting \cref{eq:he_theta_derivative} into \cref{eq:momflux_derivation_thirdstep}, dotting by \(\grad y\), and making use of
\begin{equation}
	\frac{1}{2\pi}\int_0^{2\pi} \rmd \theta \: w_i w_j = \frac{w_\perp^2}{2}(\delta_{ij} - \hat{z}_i\hat{z}_j) + w_\parallel^2 \hat{z}_i\hat{z}_j,
\end{equation}
we find
\begin{equation}
	\intw (\vec{w} \bcdot \grad y) h_e = \intw \frac{w_\perp^2}{2\Omega_e} \frac{\partial h_e}{\partial x}.
	\label{eq:momflux_derivation_laststep}
\end{equation}
Then, using \(\vthe/\Omega_e = -\rhoe\) and \cref{eq:momflux_derivation_laststep}, \cref{eq:momflux_derivation_firststep} reduces to
\begin{align}
	\Pi = \frac{\rhoe^2m_e}{4} \int \frac{\rmd^3 \vec{r}}{V} \intw \frac{\partial \phinorm}{\partial y} w_\perp^2 \frac{\partial h_e}{\partial x}.
	\label{eq:momflux_derivation_laststep_butreallyitsthelastonethistime}
\end{align}
Finally, to lowest order in \(\kperp \rhoe \ll 1\), \(h_e\) is given by \citep{adkins23}
\begin{equation}
	h_e = \left[\frac{\delta n_e}{n_{e}} - \phinorm + \frac{\delta T_e}{T_{e}} \left(\frac{w^2}{\vthe^2} - \frac{3}{2}\right)\right] \Fe,
	\label{eq:he_expression}
\end{equation}
where the density perturbation is given by \cref{eq:qn_etg}. Therefore, \cref{eq:momflux_derivation_laststep_butreallyitsthelastonethistime} becomes
\begin{equation}
	\Pi =-\frac{n_{e} T_{e} \rhoe^2}{2} \int \frac{\rmd^3 \vec{r}}{V} \frac{\partial \phinorm}{\partial y} \frac{\partial}{\partial x} \left[\left(1 + \frac{Z T_{e}}{T_{i}}\right) \phinorm - \frac{\delta T_e}{T_{e}}\right]
\end{equation}
to lowest order in \(\kperp \rhoe \ll 1\), which is exactly \cref{eq:momflux_etg}.

\bibliography{bib.bib}
\bibliographystyle{jpp}    
	
\end{document}